\documentclass[11pt]{article}
\usepackage{graphicx}
\usepackage{feynmf}
\usepackage{epsfig}
\usepackage{here}
\usepackage{wrapfig}
\usepackage{psfrag}
\usepackage{color}
\usepackage{lscape}
\usepackage{rotating}

\newcommand{\BABARPubYear}    {03}

\newcommand{\BABARConfNumber} {013}
\newcommand{\SLACPubNumber} {10067}

\newcommand{\smallrule}{\rule[-2.0mm]{0.0cm}{0.9cm}}

\input babarsym

\def\mmx {\ensuremath{\langle M_X \rangle}}
\def\mmxs {\ensuremath{\langle M_X^2 \rangle}}

\def\mmxn {\ensuremath{\langle M_X^n \rangle~}}
\def\mmxni {\ensuremath{\langle M_{X,i}^n \rangle~}}

\def\psm {\ensuremath{p^*_{min}}}

\def\breco {\ensuremath{B_{reco}}}
\def\vcb {\ensuremath{|V_{cb}|}}
\def\vub {\ensuremath{|V_{ub}|~}}

\long\def\inst#1{\par\nobreak\kern 4pt\nobreak
    {\it #1}\par\vskip 10pt plus 3pt minus 3pt}

\begin{document}
{\pagestyle{empty}

\begin{flushright}
\babar-CONF-\BABARPubYear/\BABARConfNumber \\
SLAC-PUB-\SLACPubNumber \\
July 2003 \\
\end{flushright}

\begin{center}
\end{center}

\par\vskip 2cm

\begin{center}

\Large \bf Measurement of the First and Second Moments of the Hadronic Mass Distribution in Semileptonic \mbox{\boldmath$ B$} Decays  \\
\end{center}
\bigskip

\begin{center}
\large The \babar\ Collaboration\\
\mbox{ }\\
\today
\end{center}
\bigskip \bigskip

\begin{center}
\large \bf Abstract
\end{center}
We  report a preliminary measurement of the first and second moments of the
hadronic mass distributions in $B\rightarrow X_c \ell \nu$ decays.
The measurements are based on $\FourS \to \BB$ events where the hadronic decay 
of one of the $B$ mesons is fully reconstructed and a charged lepton from the 
decay of the other $B$ meson is identified.
The moments are presented for threshold lepton momenta ranging from 0.9 to 1.6 \gev.  
From the \mmxs\ moments we determine the non-perturbative Heavy Quark Expansion (HQE) 
parameters, $\bar{\Lambda}$ and $\lambda_1$. We combine the measured moments \mmxs\ 
with earlier \babar\ measurements of the semileptonic branching ratios and \B\ 
lifetimes and perform a simultaneous fit to the HQE for the moments obtained for 
different threshold lepton momenta and the semileptonic decay width. This fit results 
in an improved value for the CKM matrix element \vcb.

\vfill
\begin{center}
Submitted at the International Europhysics Conference On High-Energy Physics (HEP 2003)  
,\\ 
17 -- 23 July 2003, Aachen, Germany
\end{center}

\vspace{1.0cm}
\begin{center}
{\em Stanford Linear Accelerator Center, Stanford University, 
Stanford, CA 94309} \\ \vspace{0.1cm}\hrule\vspace{0.1cm}
Work supported in part by Department of Energy contract DE-AC03-76SF00515.
\end{center}

\newpage
} 

\begin{center}
\small

The \babar\ Collaboration,
\bigskip

%
B.~Aubert,
R.~Barate,
D.~Boutigny,
J.-M.~Gaillard,
A.~Hicheur,
Y.~Karyotakis,
J.~P.~Lees,
P.~Robbe,
V.~Tisserand,
A.~Zghiche
\inst{Laboratoire de Physique des Particules, F-74941 Annecy-le-Vieux, France }
% ...

\end{center}\newpage

\section{Introduction}
\label{sec:Introduction}

The heavy quark limit in QCD has become a very useful tool for relating inclusive $B$-decay properties, like the semileptonic branching fraction
and moments of mass and lepton momentum distributions, 
to the charged current couplings, \vcb\ and \vub. 

In this paper we report a measurement of the moments of the invariant mass distributions, \mmx\ and \mmxs, of the hadronic system recoiling against the charged lepton and the neutrino in semileptonic $B$ decays. The moments are presented for threshold momenta of the charged leptons, ranging from 0.9 \gev
to 1.6 \gev.   
This momentum range retains the sensitivity to the spectrum of  
hadronic final states, including higher mass resonant charm mesons and non-resonant $D^{(*)}\pi$ states.

A preliminary measurement of the threshold momentum dependence of \mmxs\ was presented last summer~\cite{Aubert:2002pm}. This earlier analysis followed a procedure first introduced by the CLEO Collaboration~\cite{Cronin-Hennessy:2001fk} in which the moments were derived from the branching fractions and average hadron mass distribution of various charm states.  For this purpose, the mass distributions were decomposed into contributions from various charm meson resonant and non-resonant states.  A limitation of this approach was that it required as input the knowledge of the mass distribution for the individual charm states, and this information as well as the branching ratios remain uncertain for the higher mass states.

The new analysis presented here is much less dependent on this input. 
The moments are extracted directly from the measured $M_X$ and $M_X^2$ distributions, taking into account corrections for the mass scale, the detection efficiency, and small residual backgrounds.

Theoretical calculations of the second moment 
$\langle M_X^2 \rangle$ have been carried out 
\cite{Falk:1998jq} using an Operator Product Expansion (OPE) in powers of the strong coupling constant $\alpha_s(m_b)$ and in inverse powers of the $B$ meson mass up to order $\alpha^2_s \beta_0$ and $1/m_B^3$. (Here 
$\beta_0=(33-2n_f)/3=25/3$ is the leading order QCD $\beta$ function, and $n_f$ is the number of quark flavors accessible at this energy, i.e. $n_f = 4$.)
These expansions contain the non-perturbative parameters $\bar{\Lambda}$ (${\cal O}(1/m_B)$), $\lambda_1$, and $\lambda_2$ (${\cal O}(1/m^2_B)$). 
The moment measurements presented here extend over a large region of phase space and thus we can use them to extract  $\bar{\Lambda}$ and $\lambda_1$  and  to test the validity of the assumptions made in the OPE.

By combining measurements of \mmxs\ for seven values of the momentum threshold
with earlier \babar\ measurements of the semileptonic branching ratios and \B\ lifetimes we can further constrain the parameter $\lambda_1$ and the $b$ quark mass, $m_b^{1S}$, in the $1S$ scheme~\cite{Bauer:2002sh} (where parameters like the quark mass are defined by reference to the mass of the $\Upsilon (1S)$ state) and derive an improved measurement of the CKM matrix element \vcb.  
Furthermore,  we use the published data on moments of hadron mass-squared, lepton energy and photon energy distribution in semileptonic and rare
radiative $B$ decays,  to test the consistency of the
underlying theoretical framework.

\section{The \babar\ Detector and Data Set}
\label{sec:babar}

The measurements presented here are based on a sample of about 89 million \BB\ pairs
collected  at the \FourS\ resonance by the \babar\
detector~\cite{Aubert:2001tu} at the
PEP-II  asymmetric-energy  $e^+e^-$ storage ring operating at SLAC.
 The detector consists of a five-layer silicon vertex tracker (SVT), a 40-layer drift chamber
(DCH), a detector of internally reflected Cherenkov light (DIRC), and an
electromagnetic calorimeter (EMC) assembled from 6580 CsI(Tl) crystals, all embedded in a solenoidal magnetic field of 1.5 T and surrounded by an instrumented flux return (IFR).

\section{Analysis Method}
\label{sec:Analysis}

This measurement of the hadronic mass distribution in semileptonic \B\ decays exploits the very large sample of \B\ mesons recorded by the \babar\ detector.  The analysis relies on \BB\ events in which one \B\ meson decays hadronically and is fully reconstructed ($B_{reco}$)  and the semileptonic decay of the other $B$ meson ($B_{recoil}$) is identified by the presence of an electron or muon. While this approach 
results in a low overall event selection efficiency, it allows for the determination of the momentum, charge,
 and flavor of the $B_{reco}$ meson. 
Furthermore, the remaining tracks and photons in the event can be used to reconstruct the hadronic mass $M_X$ of the semileptonic decay. The determination of $M_X$ is substantially improved by application of a kinematic fit to the full event.

\subsection{Selection of Fully Reconstructed Hadronic \mbox{\boldmath$ B$} Decays, \mbox{\boldmath$ B \ra D Y$}}

To reconstruct a large sample of $B$ mesons, hadronic decays of the type $B_{reco}
\rightarrow  \Db^{(*)} Y^{\pm}$  are  selected.  Here, $D^{(*)}$ refers to a charm meson and $Y^{\pm}$ consists of hadrons with a total charge of $\pm 1$, composed
of $n_1\pi^{\pm}\, n_2K^{\pm}\, n_3\KS\,  n_4\piz$ with $n_1 + n_2 \leq
5$,  $n_3  \leq  2$,  and  $n_4  \leq  2$.   We  reconstruct  $D^{*-}\ra
\Dzb\pi^-$ and $\Dstarzb \ra
\Dzb\piz, \Dzb\gamma$ and the decays $D^-\ra K^+\pi^-\pi^-$, $K^+\pi^-\pi^-\piz$, $\KS\pi^-$,
$\KS\pi^-\piz$, $\KS\pi^-\pi^-\pi^+$; and $\Dzb\ra K^+\pi^-$,
$K^+\pi^-\piz$, $K^+\pi^-\pi^-\pi^+$,  $\KS\pi^+\pi^-$. 

The kinematic consistency of the $B_{reco}$ candidates 
is checked with two variables,
the beam energy-substituted mass $\mes = \sqrt{s/4 -
\vec{p}^{\,2}_B}$ and the energy difference 
$\Delta E = E_B - \sqrt{s}/2$. Here $\sqrt{s}$ is the total
energy in the \FourS\ center-of-mass frame, and $\vec{p}_B$ and $E_B$
denote the momentum and energy of the $B_{reco}$ candidate in the same
frame.  
We require $\Delta E$ to be within three standard deviations from zero for each decay mode.

For a given \breco\ decay mode, the purity
is estimated as the fraction of signal events with \mes$>
5.27$\gev.  For this analysis we only use decay modes for which the purity
exceeds 40\%. 
In events with more than one
reconstructed \breco\ decay, we select the decay mode with the highest purity. For the case of two candidates being reconstructed in the same mode, the one with $\Delta E$ closest to zero is selected.

\subsection{Selection of Semileptonic Decays, \mbox{\boldmath$B \ra X \ell \nu$}}

Semileptonic $B$ decays are identified by the presence of one  and only one electron or muon above a minimum momentum $p^*_{min}$  measured in the rest frame of the $B_{recoil}$ meson recoiling against the $B_{reco}$. The lepton momentum threshold $p^*_{min}$ is varied in the range of $0.9 -1.6 \gev$.
The higher the lepton momentum, the lower are the residual backgrounds from secondary decays of charm particles, $\tau^{\pm}$ leptons, and from misidentified hadrons. Measurements with lower lepton momenta are more sensitive to the production of higher mass charm mesons, in the form of both resonant and non-resonant states.

To improve the background rejection, we impose a condition on $Q_{\ell}$, the charge of the lepton from $B_{recoil}$, and $Q_b$, the charge of the $b$-quark of $B_{reco}$, namely $Q_b Q_{\ell} < 0$.  This condition is fulfilled for primary leptons, except for \BzBzb\ events in which flavor mixing has occured.

Electrons are identified using a likelihood-based algorithm, combining the track momentum with the energy, position, and shape of the shower measured in the EMC, the Cherenkov angle and the number of photons measured in the DIRC, and the specific energy loss in the DCH.
The efficiency of the electron selection has been measured with radiative Bhabha events and has been corrected for the higher multiplicity of \BB events using Monte Carlo (MC) simulations. 

Muons are identified using a cut-based selection for minimum ionizing tracks, relying on information from the finely segmented instrumented flux return of the magnet. The number of interaction lengths traversed by the track, the spatial width of the observed signals, and the match between the IFR hits and the extrapolated charged track are used in the selection. 
The muon identification efficiency has been measured with $\mu^+\mu^-
(\gamma)$ events and two-photon production of $\mu^+\mu^-$ pairs. 

The misidentification probabilities for pions,  kaons, and protons  have been extracted from  selected data control samples. They vary between 
0.05\% and 0.1\% (1.0\% and 3\%) for the electron (muon) selection.

To further reduce backgrounds we require the total charge of the event to be $|Q_{tot}|= |Q_{Breco} + Q_{Brecoil}| \leq 1$.
We explicitly allow for a charge imbalance to reduce the dependence on the exact modeling of charged particle tracking in the Monte Carlo simulation, especially at low momenta, and the production of tracks from photon conversions.

\subsection{Reconstruction of the Hadronic Mass \mbox{\boldmath$M_X$}} 
\label{subsec:kfit}

The hadronic mass $M_X$ in the decay $B \ra X \ell \nu$ is reconstructed from charged tracks ($\pi^{\pm}$ and $K^{\pm}$) and photons that are not associated with the $B_{reco}$ candidate or identified as leptons. Specifically, we select charged tracks 
with a minimum transverse momentum $p_t > 0.1 \gev$ and at least 12 DCH hits. For photons we require an energy sum greater than $0.05 \gev$ deposited in three or more crystals.
To suppress background from hadrons interacting in the calorimeter we require a minimum angular separation between the center of the shower and the nearest track impact point of 200\mrad and a shower shape consistent with a photon (lateral moment $LAT <0.5$, defined in reference~\cite{Aubert:2001tu} ). 

The measured four-momentum $P_X$ of the hadron $X$ can be written as
\begin{equation}
P_X = \sum_{i=1}^{N_{ch}} P^{ch}_i  
      + \sum_{j=1}^{N_{\gamma}} P^{\gamma}_j 
\end{equation}

\noindent
where $P$ are four-momenta and the superscripts $ch$ and $\gamma$ 
refer to the selected charged tracks and photons, respectively. Depending on particle identification the charged tracks are assigned either the $K^{\pm}$ or $\pi^{\pm}$ mass.

\begin{figure}[t]
\begin{center}
\begin{picture}(15.,6.)
\put(-1.0,0.0){\mbox{\epsfig{file=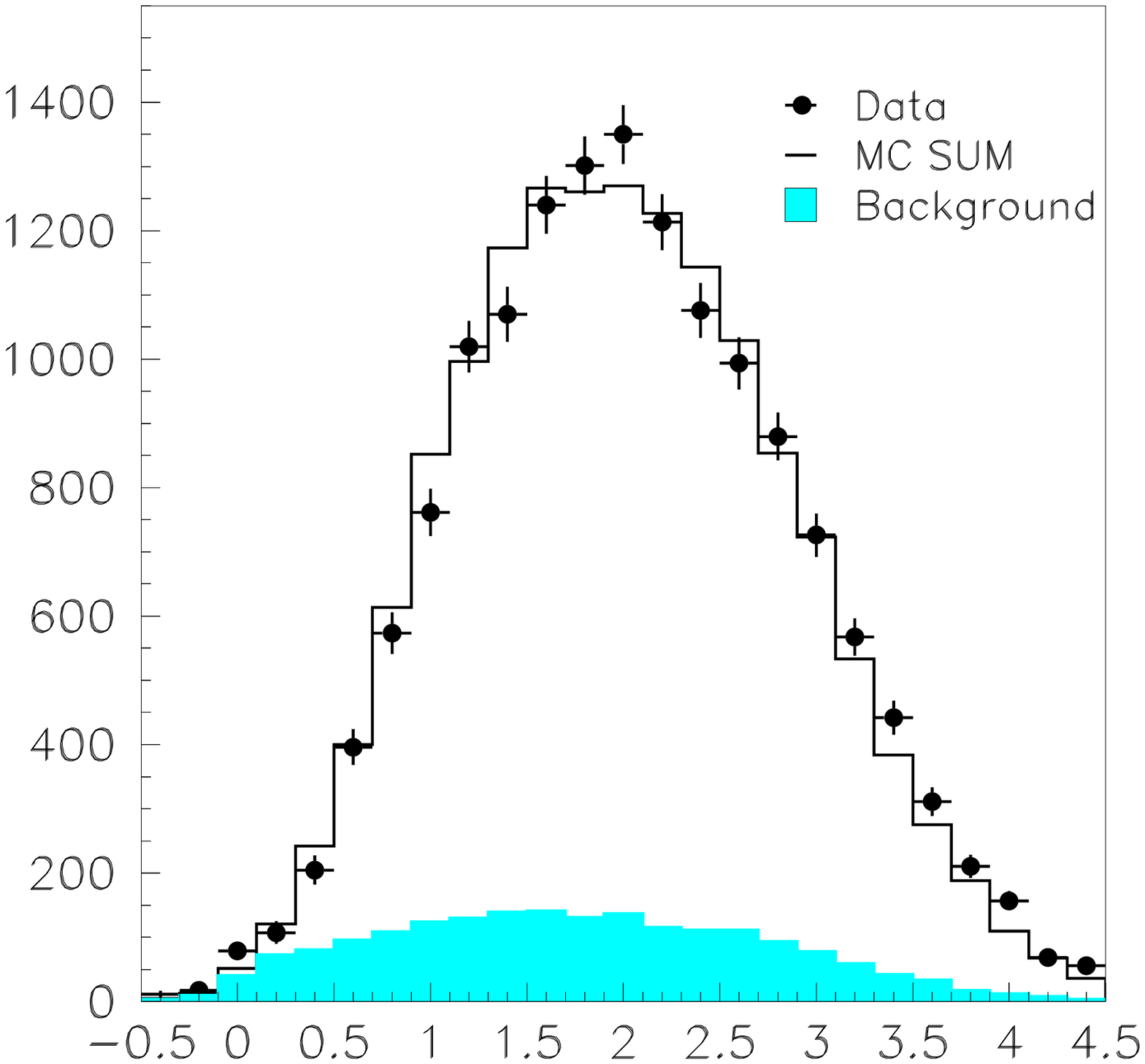,scale=0.28}}}
\put(4.9,0.0){\mbox{\epsfig{file=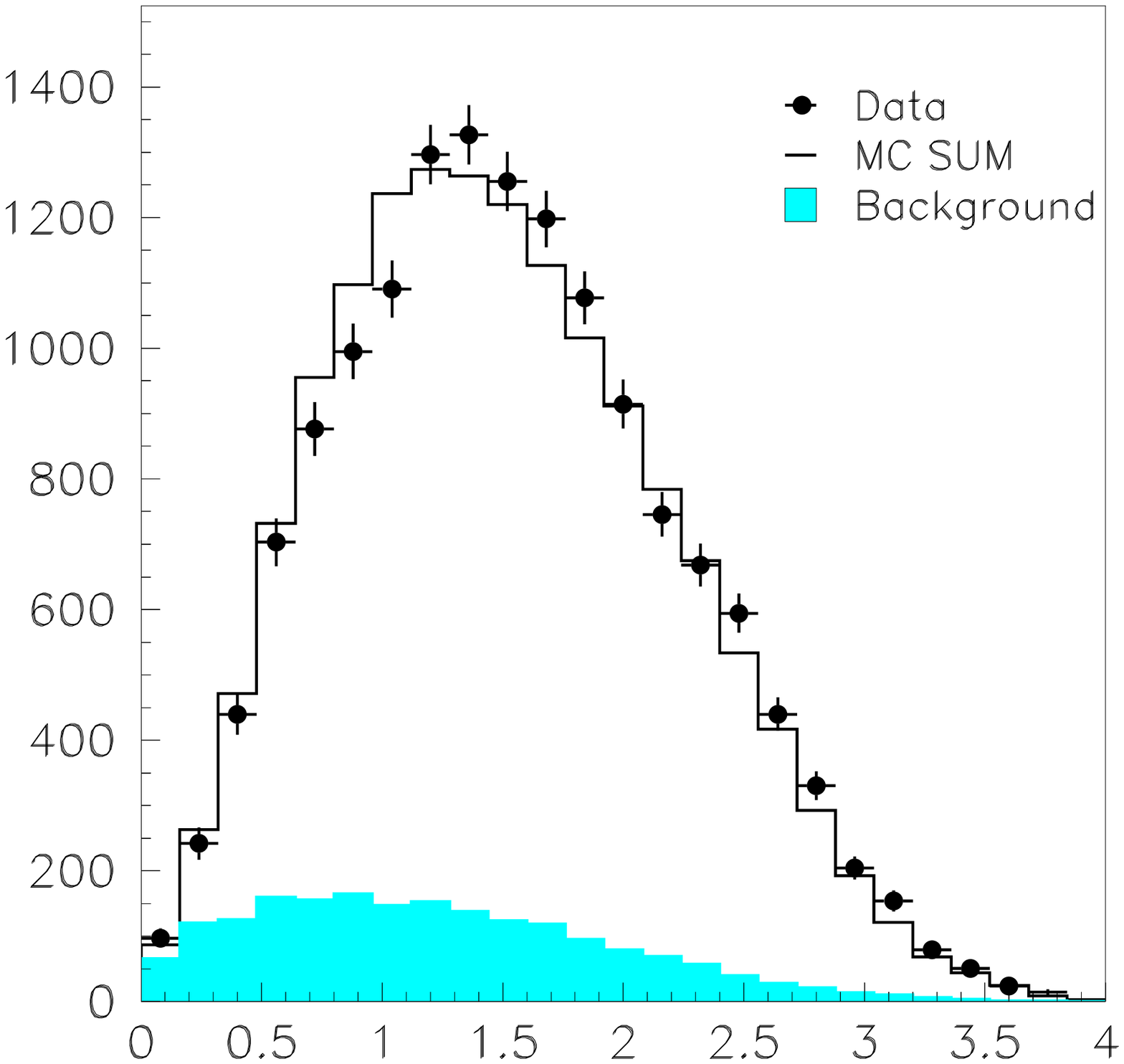,scale=0.28}}}
\put(10.8,0.0){\mbox{\epsfig{file=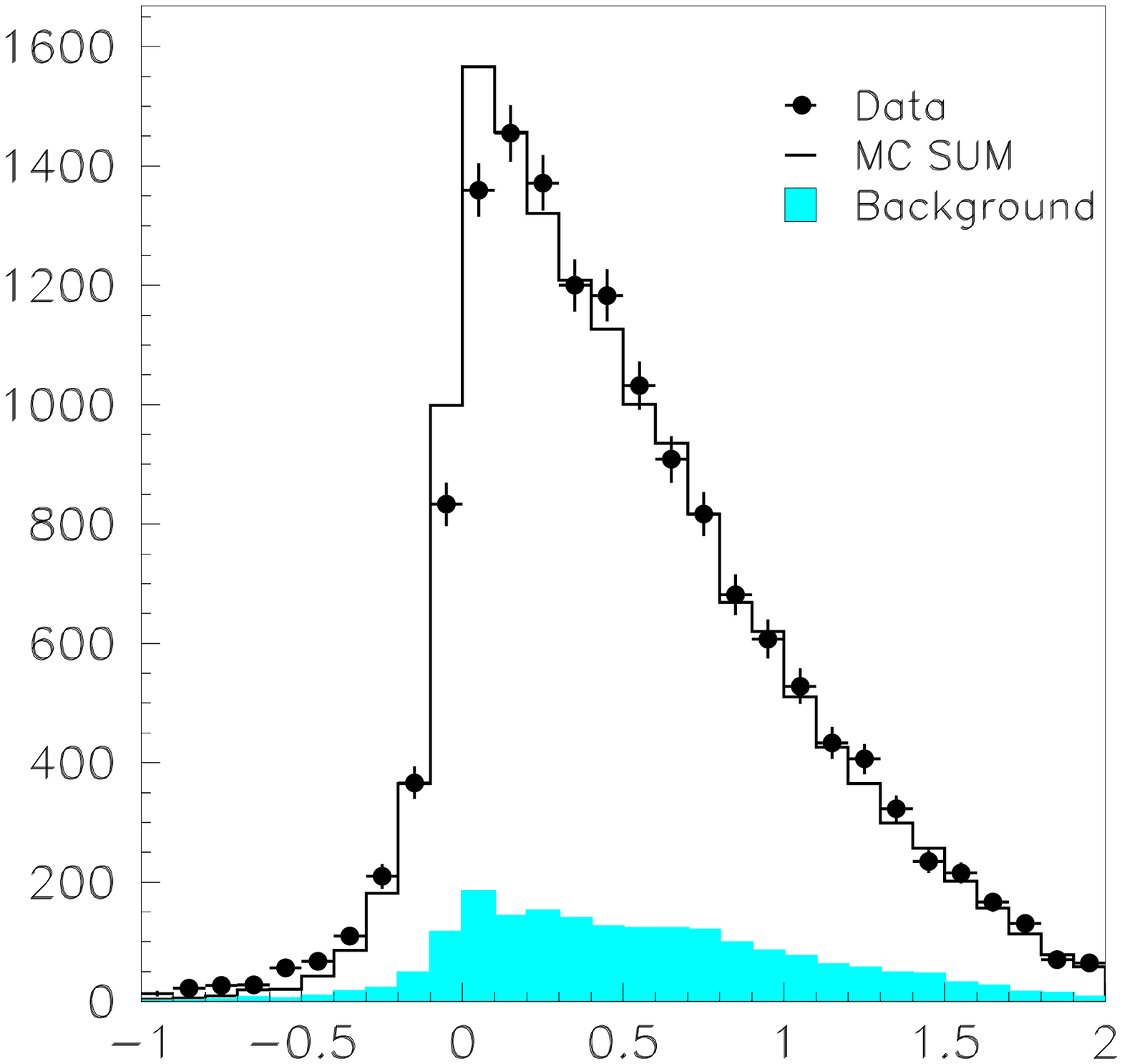,scale=0.28}}}
\put(2.1,-0.2){$E_{miss} (\gev)$}
\put(7.8,-0.2){$|\vec {p}_{miss}| (\gev)$}
\put(12.3,-0.2){$E_{miss}-|\vec {p}_{miss}| (\gev)$}
\put(-1.3,3.0){\mbox{\begin{turn}{90}{$N/ 200 \mev$}\end{turn}}}
\put(4.5,3.0){\mbox{\begin{turn}{90}{$N/ 160 \mev$}\end{turn}}}
\put(10.5,3.0){\mbox{\begin{turn}{90}{$N/ 100 \mev$}\end{turn}}}
\end{picture}   
\end{center}
\caption{\em Data - MC comparison of missing energy $E_{miss}$ (left), missing momentum $|\vec {p}_{miss}|$ (middle) and $E_{miss}-|\vec {p}_{miss}|$ (right) for $\psm =0.9~\gev$.  The MC predictions for the total distribution as well as the residual background are indicated. 
\label{fig:datamccomp} }
\end{figure}

The neutrino four-momentum $P_{\nu}$ is estimated from the
missing momentum four-vector,
$P_{miss} = P_{\Upsilon(4S)}-P_{B_{reco}}-P_X - P_\ell$, 
where all momenta are measured in the laboratory frame.
The measured missing momentum four-vector
is an important indicator of the quality of  the reconstruction of the total recoil system. 
Any particle that is undetected or poorly measured, any sizable energy deposition due to charged hadron interactions or beam-generated background will impact the measurement of $M_X$ and $M_{miss}^2 = P_{miss}^2$.  To reduce the impact of these experimental effects, we impose the following criteria:  $E_{miss} > 0.5 \gev$, $|\vec {p}_{miss}| > 0.5 \gev$, and $|E_{miss} - |\vec {p}_{miss}|| < 0.5 \gev$.
In Figure~\ref{fig:datamccomp} the $E_{miss}$, $|\vec {p}_{miss}|$, and  $E_{miss} - |\vec {p}_{miss}|$ distributions are shown and compared to MC simulations, normalized to the total number of events.  
The agreement reflects our 
understanding of the detector performance and improvements in the suppression of track and photon background which have been studied extensively. It should be noted that the branching fractions for the individual decay modes used in the MC can also contribute to differences between data and MC. However, the extraction method for the moment measurements is not sensitive to branching fraction variations.

To further improve the resolution on the measurement of the hadronic mass $M_X$, we exploit the available kinematic information from the full event, namely the $B_{reco}$ and the $B_{recoil}$ candidate, by performing a kinematic fit with two constraints that imposes four-momentum conservation, the equality of the masses of the two $B$ mesons, $M_{recoil}=M_{reco}$, and forces $M_{miss}^2 = 0$. The fit takes into account event-by-event measurement errors for all individual particles and the four-vector of the measured missing momentum.  The resulting $M_X$ resolution is 350 \mev.

\subsection{Subtraction of \mbox{\boldmath$B_{reco}$} Background} 
\label{subsec:mesfit}

\begin{figure}[t]
\begin{center}
\begin{picture}(15.,7.5)
\put(2.0,0.0){\mbox{\epsfig{file=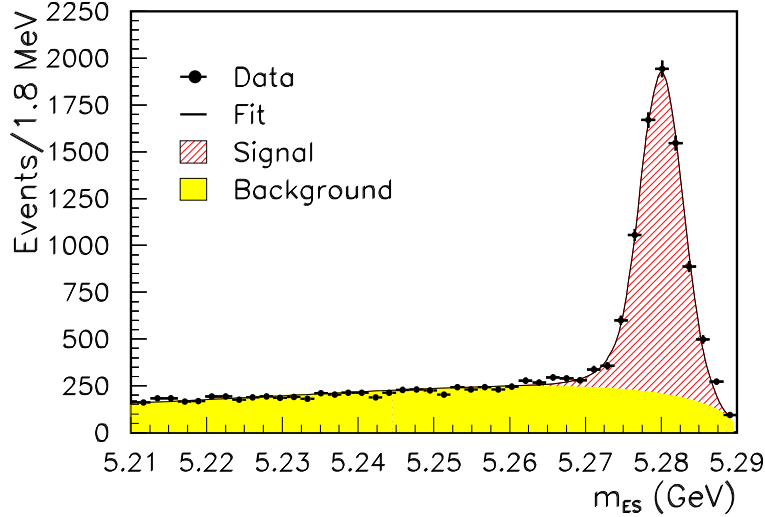,scale=0.7}}}
\end{picture}   
\end{center}
\vspace{-1.0cm}
\caption{\em The $m_{ES}$ distribution for $B_{reco}$ candidates in events for which the $B_{recoil}$ meets all selection criteria for a semileptonic decay. The lepton momentum threshold is $p^*_{min}=0.9 \gev$.
\label{fig:mes} }
\end{figure}

Figure ~\ref{fig:mes} shows the \mes distribution for the $B_{reco}$ candidates for all events for which the $B_{recoil}$ meson meets all selection criteria for a semileptonic decay. 
To estimate the combinatorial background in the signal region 
($\mes > 5.27 \gev$), the measured \mes distribution is fitted to a sum of a background function (introduced by the ARGUS Collaboration~\cite{Albrecht:1987nr}) and a signal function (first used by the Crystal Ball Collaboration~\cite{Skwarnicki:1986xj}).  
For $\mes > 5.27 \gev$, we find a total of $7114$ signal events above a background of $2102$ events.  
In the analysis, the \breco\ background subtraction is carried out for several regions in $M_X$ to account for changes in the signal-to-background ratio as a function of $M_X$.
\unitlength1.0cm 

\section{Determination of the Hadronic Mass Moments}
\label{sec:Moments}

\begin{figure}[H]
\begin{center}
\begin{picture}(15.,14.)
\put(0.0,7.5){\mbox{\epsfig{file=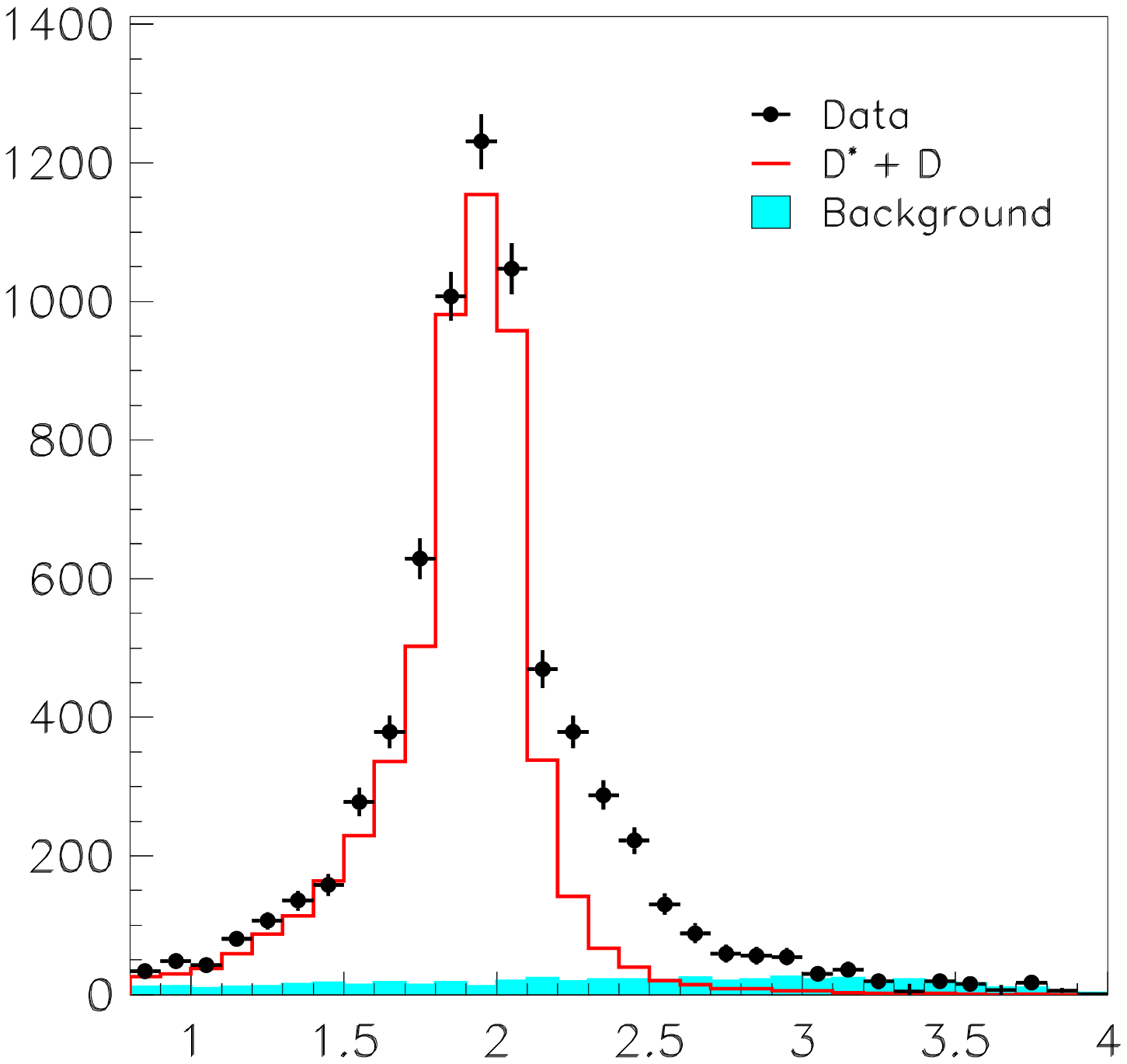, scale=.45}}}
\put(7.7,7.5){\mbox{\epsfig{file=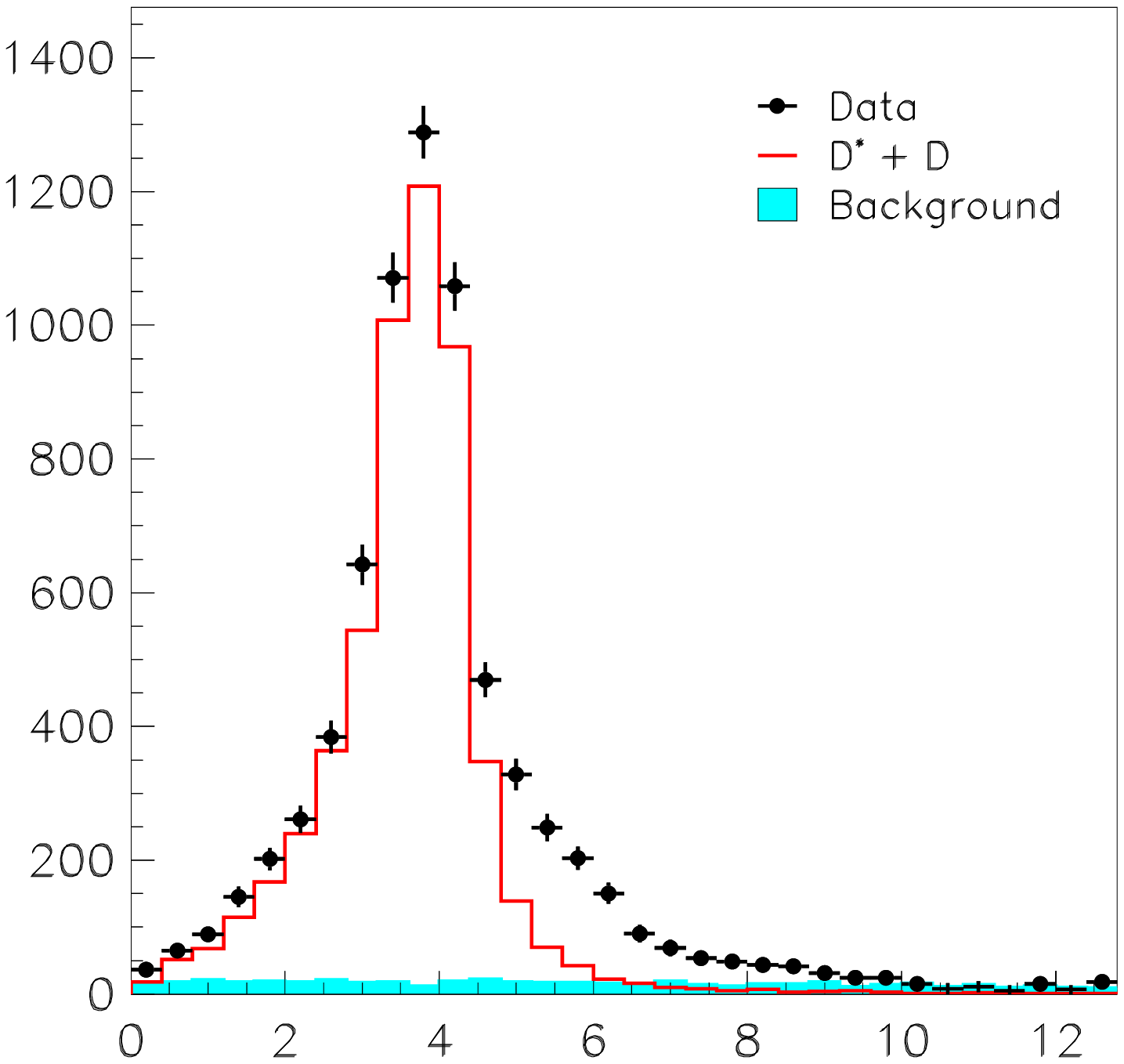, scale=.45}}}
\put(0.0,0.0){\mbox{\epsfig{file=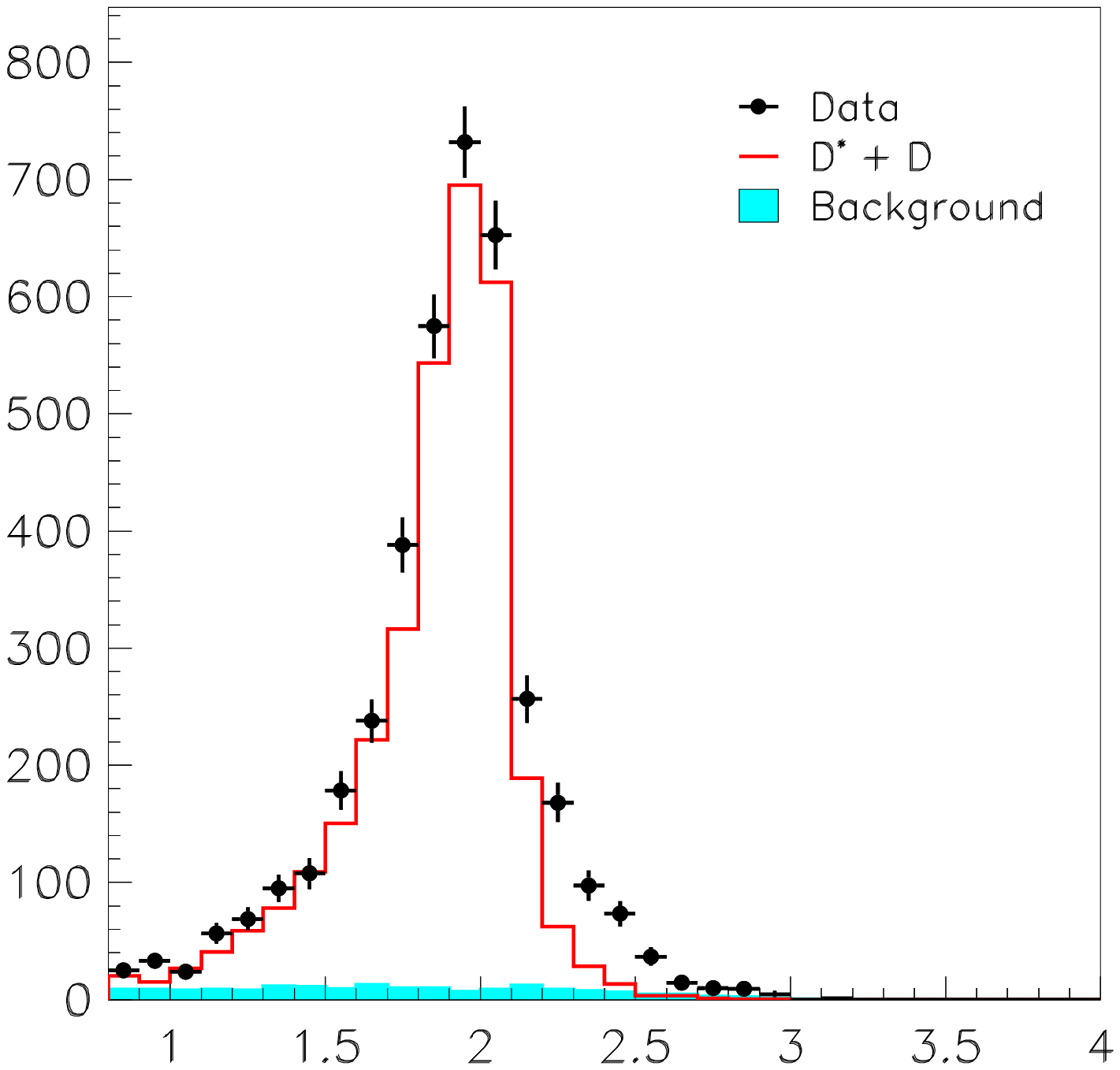, scale=.45}}}
\put(7.7,0.0){\mbox{\epsfig{file=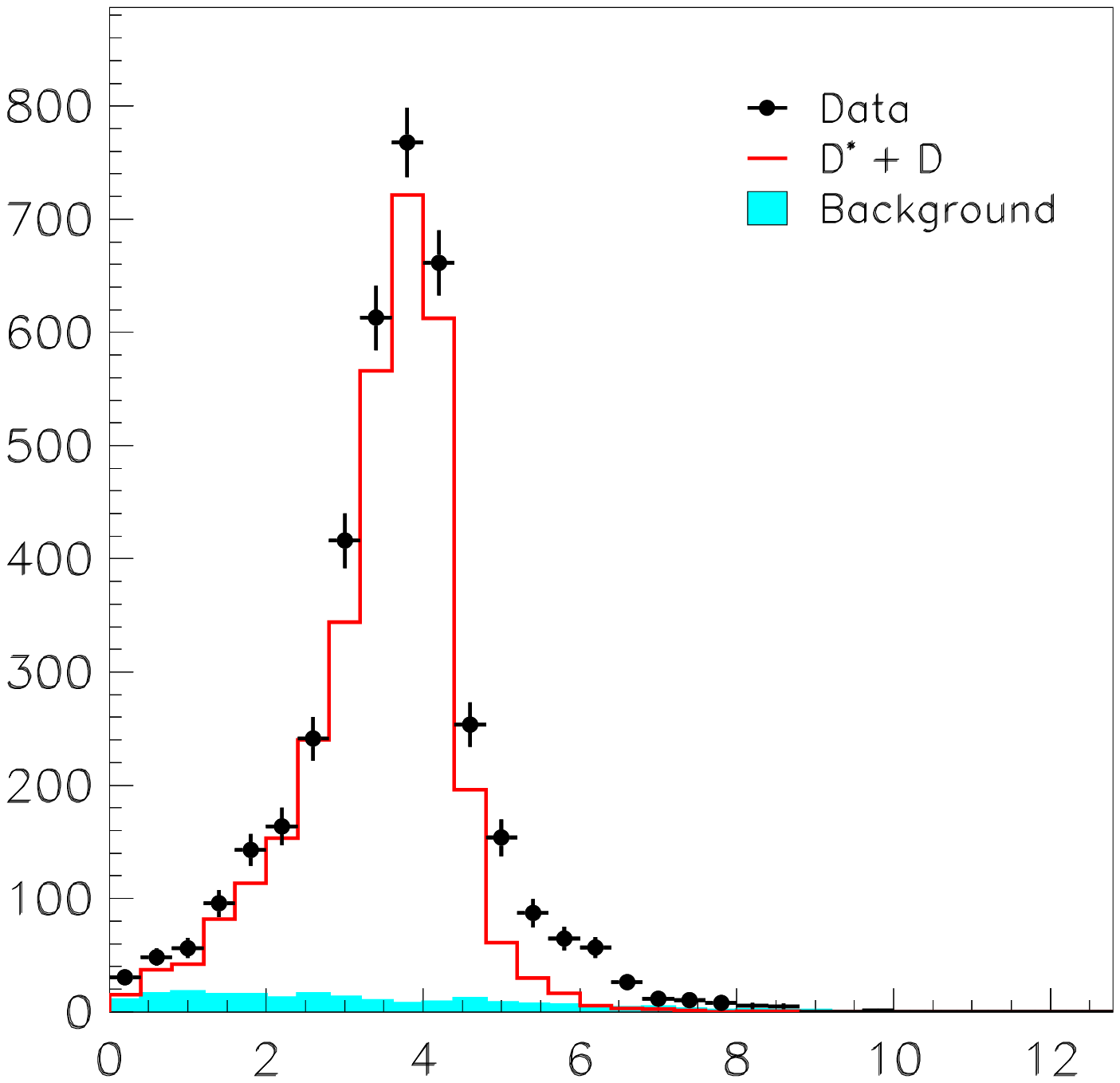, scale=.45}}}
\put(-0.4,4.5){\mbox{\begin{turn}{90}$N/ 100 \mev$  \end{turn}}}
\put(-0.4,12.1){\mbox{\begin{turn}{90}$N/ 100 \mev$  \end{turn}}}
\put(7.3,4.3){\mbox{\begin{turn}{90}$N/ 400 \mev^2$  \end{turn}}}
\put(7.3,11.9){\mbox{\begin{turn}{90}$N/ 400 \mev^2$  \end{turn}}}
\put(4.8,0.0){$M_X (\gev)$ }
\put(12.3,0.0){$M_X^2 (\gev^2)$}
\put(4.7,7.4){$M_X (\gev)$}
\put(12.3,7.4){$M_X^2 (\gev^2)$}

\end{picture}   
\end{center}
\caption{\em \label{mx_dist} $M_X$ (left) and $M_X^2$ (right) distributions after subtraction of the \breco\ background, with $\psm = 0.9~\gev$ (top) and $\psm = 1.5~\gev$ (bottom).
The dominant contributions from $D$ and $D^*$ mesons (histogram) and the residual background (shaded area) as estimated from MC are indicated.  } 
\end{figure}

The distributions in $M_X$ and $M_X^2$, corrected for combinatorial $B_{reco}$ background, are shown in Figure~\ref{mx_dist} for two different lepton threshold momenta. 
The dominant contributions are from the lowest mass mesons, 
($D^+$, $D^0$) and ($D^{*0}$, $D^{*+}$), but there are clear indications
of the higher mass charm meson states. 
The residual background, estimated from MC simulation, is primarily due to  hadron misidentification and non-prompt leptons from semileptonic decays of $D^{(*)}$ and $D_s$ mesons, either from \BzBzb\ mixed events or produced in $W^-\to \cbar s$ fragmentation.

To extract the first moments from these distributions, i.e., the unbiased mean values \mmx\ and \mmxs, we need to correct for effects that can distort these distributions and thereby affect the moments.
Specifically, we need to calibrate the measurement of $M_X$ and $M_X^2$ by relating the observed values to the generated values for MC simulated distributions.  Furthermore, we need to reliably estimate and subtract the contribution to the moments from residual backgrounds and then correct the result for the detection efficiency.  Thus we can express the true hadronic mass moment $\mmxn$ as 
\begin{equation}
\label{fineq2}
    \mmxn = \frac{\mmxn^{DATA}_{calib} - f_{bg} \cdot \mmxn^{MCbg}_{calib}} 
{1-f_{bg}} \times \frac{\sum\limits_{i=1}^{10} R_{i} \mmxni^{MCtrue}}                                                                         {\sum\limits_{i=1}^{10} R^{det}_{i}\mmxni_{calib}^{MCdet}}
\end{equation}
\noindent where the first term describes the subtraction of the residual background and the second term specifies the efficiency correction. Here $\mmxn^{sample}_{calib}$ (with $sample=DATA, MCbg, MCtrue, MCdet$) refer to the mass moments (calibrated and corrected for $B_{reco}$ background) extracted from samples of data, of MC-simulated background, and of ten MC-generated  and -detected exclusive semileptonic decays, $B \to X_c^i \ell \nu$.  These decay modes, identified by the index $i$,  include $D$, $D^*$, four resonant $D^{**}$ and four non-resonant $D^{(*)} \pi$ charm meson final states, summed over charged and neutral $B$ mesons. 
The coefficient $R_i$ is the relative
 fraction of a decay mode predicted by MC before detection efficiencies have been taken into account; $R_i^{det}$ are the corresponding fractions
 for detected decay modes.
We denote by $f_{bg}$ the size of the residual background relative to the data. 

To study the calibration, efficiencies, and backgrounds we rely on Monte Carlo simulations.  
Decays to $D^* \ell \nu$ are modeled by HQET-derived form factors~\cite{Duboscq:1996mv},
decays to $D \ell \nu$ and $D^{**} \ell \nu$ are simulated using the ISGW2 model~\cite{ISGW2}, while the non-resonant decays $D^{(*)} \pi \ell \nu$ are modeled according to the prescription  by Goity and Roberts~\cite{J.GoityandW.Roberts}.
The relative branching ratios are adjusted to be consistent with current measurements.

\begin{figure}[H]
  \begin{center}
\begin{picture}(15.,8.5)
   \put(-0.8,0.){\mbox{\epsfig{file=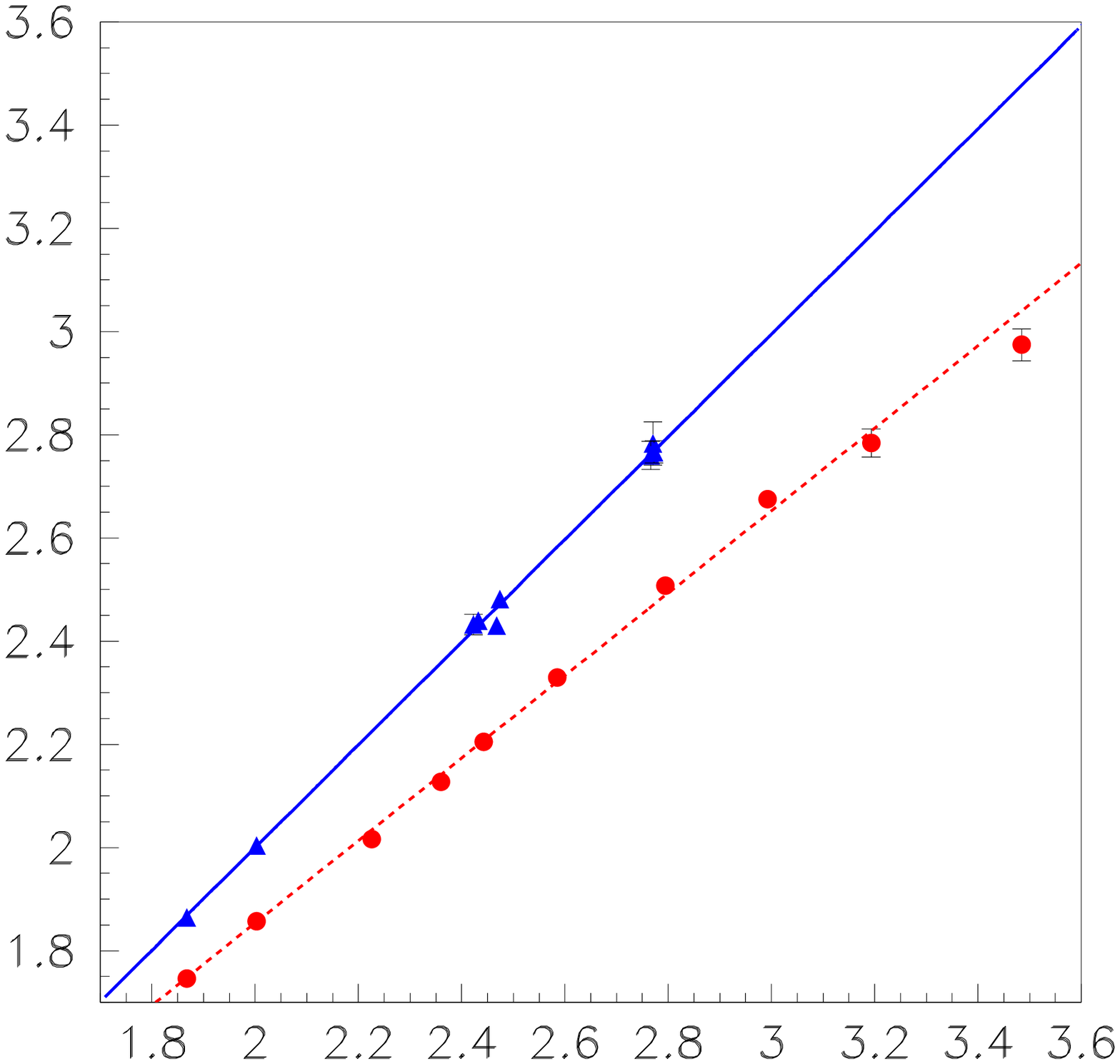,scale=0.44}}}
   \put(7.8,0.){\mbox{\epsfig{file=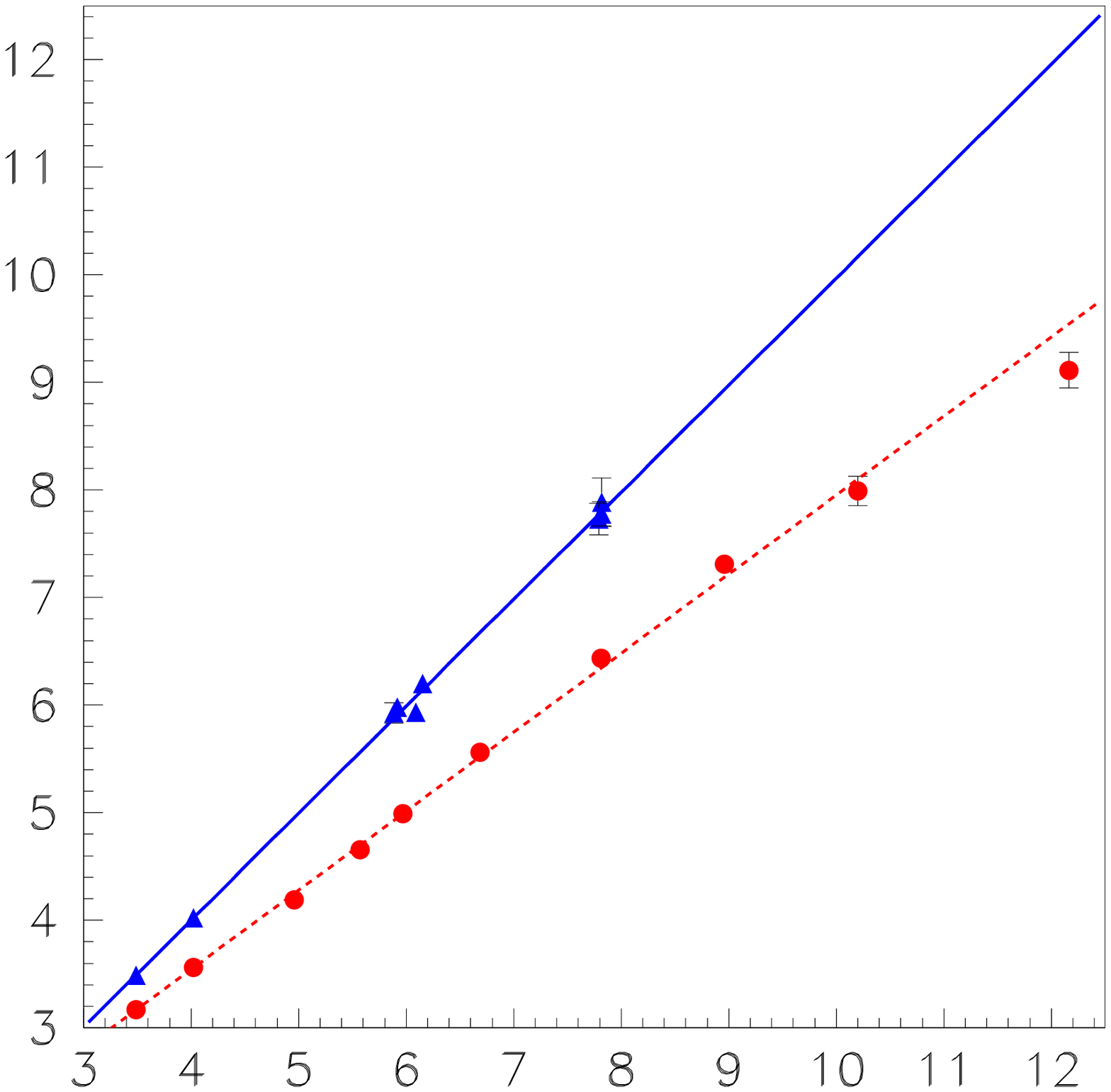,scale=0.44}}}
   \put(0.5,7.5){a)}
   \put(9.0,7.5){b)}
   \put(3.2,1.8){\mbox{\color{red} slope = $0.799\pm0.007$}}
   \put(3.2,1.3){\mbox{\textcolor{red}{offset = $0.256\pm0.016$}}}
   \put(0.5,6.8){\mbox{\color{blue}slope = $0.995\pm0.011$}}
   \put(0.5,6.3){\mbox{\color{blue}offset = $0.010\pm0.022$}}
   \put(11.6,1.8){\mbox{\color{red}slope = $0.735\pm0.007$}}
   \put(11.6,1.3){\mbox{\color{red}offset = $0.606\pm0.030$}}
   \put(9.0,6.8){\mbox{\color{blue}slope = $0.995\pm0.011$}}
   \put(9.0,6.3){\mbox{\color{blue}offset = $0.021\pm0.046$}}
   \put(3.9,0.0){$\mmx^{MCtrue} (\gev)$}
   \put(12.3,0.0){$\mmxs^{MCtrue} (\gev^2)$}
   \put(-1.1,4.8){\mbox{\begin{turn}{90}$\mmx^{MCdet} (\gev)$   \end{turn}}}
   \put(7.5,4.6){\mbox{\begin{turn}{90}$\mmxs^{MCdet} (\gev^2)$   \end{turn}}}
\end{picture}
  \end{center}
  \caption{\em \label{calibmx} Results of the calibration procedure  for events with a lepton threshold of $p^*_{min} = 0.9~GeV$, a) for $\langle M_X \rangle$ and b) for $\langle M_X^2 \rangle$. 
The calibration data and fit results are shown in the lower half (round data points and dashed line), the verification in the upper half (triangles and solid line) of the figures.  
}
\end{figure}

The calibration of the mass scale is performed separately for $\langle M_X \rangle$ and $\langle M_X^2 \rangle$.  The results of the procedure are illustrated in Figure~\ref{calibmx}.  
The average measured versus the average true value for bins in $M_X$ and $M_X^2$ are determined from a MC simulation of decays $B\to X_c \ell \nu$.  The relation between the measured and true quantities can be approximated by a linear function in both cases. Though the calibrations are obtained from averages in bins of $M_X$ and $M_X^2$, they can be applied on an event-by-event basis using the slope and offset of the linear functions obtained from fits. 

\unitlength1.0cm 

To verify this, the calibrated $M_X^n$ are averaged for each of the ten decay modes. 
As can be seen in the upper halves of Figures~\ref{calibmx}a and b, the calibrated moments $\mmxn_{calib}^{MCdet}$ reproduce the underlying true moments $\mmxni^{MCtrue}$ for each decay mode. 
This result indicates that there is no significant mass dependent or decay-type dependent bias that is unaccounted for. The calibration is largely independent of the decay model, and the impact of decays which are not included in the MC simulation (or which are mediated by other processes) can be treated adequately.

Detailed studies show that the slope and offset of the calibration curves vary slightly as a function of the multiplicity of the hadron system and $E_{miss} -|\vec {p}_{miss}|$. Thus instead of one universal calibration curve for all data, we split the data in three bins in multiplicity and three bins in $E_{miss} -|\vec {p}_{miss}|$, and derive a total of nine calibration curves, one for each subsample.  

Decays to higher mass final states usually correspond to higher multiplicities in the decay. Therefore, the event selection criteria are expected to have a stronger impact on high mass states. In addition, the different decay modes have different spin configurations and thus different angular distributions. The second term in Equation~\ref{fineq2} accounts for these effects. It is determined by MC simulation of the individual semileptonic decays $B \ra X_c^i \ell \nu$. 

\begin{figure}[H]
\begin{center}
\begin{picture}(15.,9.5)
\put(-0.8,0.0){\mbox{\epsfig{file=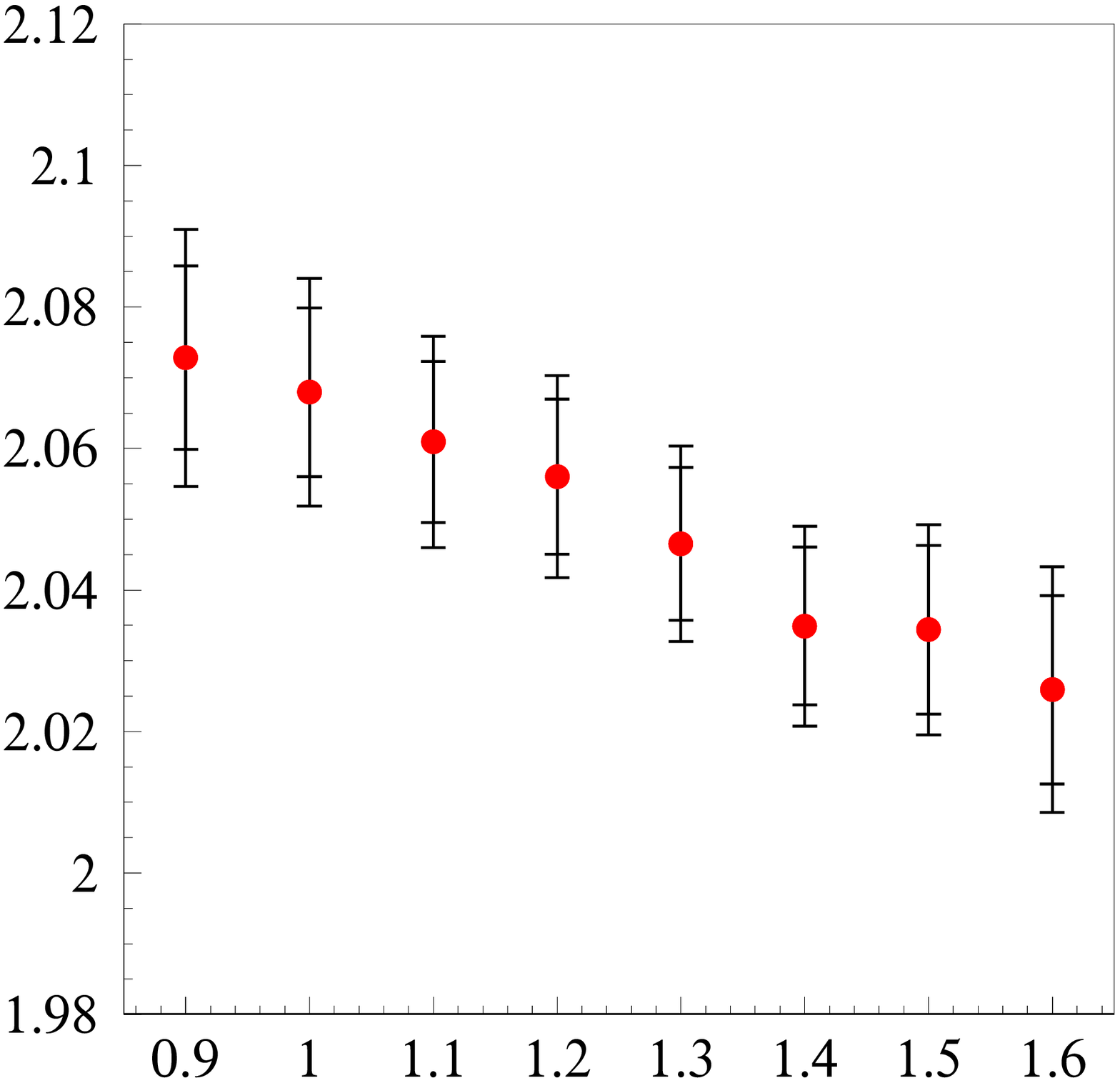, scale=.44}}}
\put(7.9,0.0){\mbox{\epsfig{file=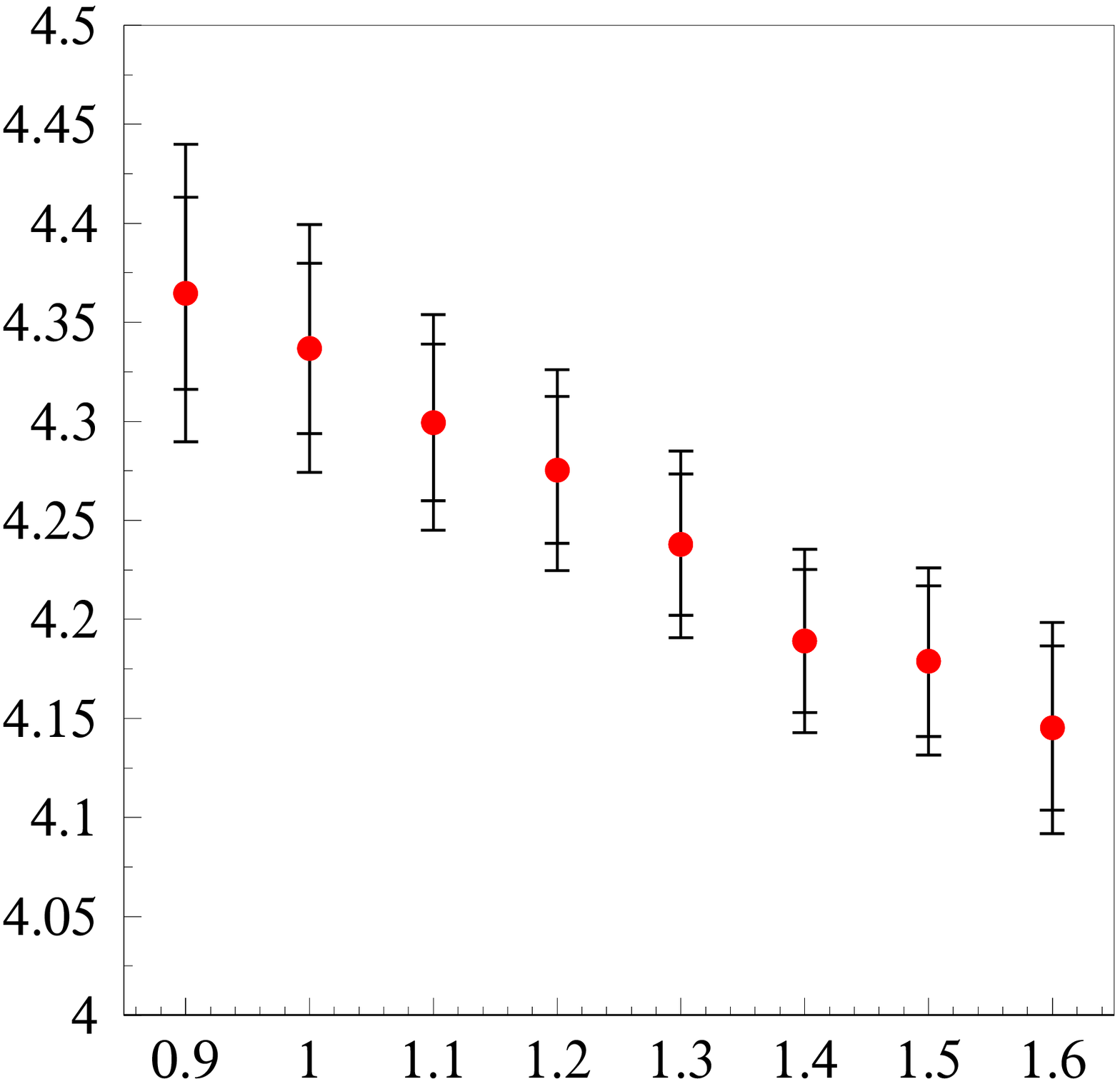, scale=.44}}}
   \put(5.3,0.0){$p^*_{min} (\gev)$}
   \put(13.9,0.0){$p^*_{min} (\gev)$}
   \put(-1.2,5.8){\mbox{\begin{turn}{90}$\mmx (\gev)$   \end{turn}}}
   \put(7.4,5.6){\mbox{\begin{turn}{90}$\mmxs (\gev^2)$   \end{turn}}}
\end{picture}   
\end{center}
\caption{\em \label{moments_1} Measured moments \mmx\ (left) and \mmxs\ (right) for different lepton threshold momenta, $p^*_{min}$. The bars indicate the statistical and the total error. The errors of the individual measurements  are highly correlated. 
}
\end{figure}

The hadronic mass moments \mmx\ and \mmxs\ obtained after background subtraction and mass calibration are presented in Figure~\ref{moments_1} as a function of \psm\ with statistical and systematic errors.
The results are also summarized in Tables~\ref{tab:mom1} and~\ref{tab:mom2}.
As expected, the values for the hadronic mass moments increase for lower \psm. 
Similarly, the systematic errors associated with background subtraction increase with lower \psm, while most of the detector related systematic errors are independent of \psm. This results in a systematic error which is of comparable size to the statistical error for most of the \psm ~range.  It should be noted that the statistical error of the measurements is almost constant as a function of \psm\ because the width of the $M_X^n$ distribution decreases as \psm\ increases.  
The correlation coefficients for the different \psm\ measurements of \mmx\ vary from $\sim 90\%$ between neighboring points to $\sim 55\%$ for the two extreme \psm\ values. These
correlation coefficients are slightly smaller for \mmxs.

\begin{table}[ht]
\begin{center}
\caption[]{\label{tab:mom1} {\it Preliminary results for \mmx\ for different values of $p^*_{min}$, with statistical and total systematic errors. The last four columns show separately the four dominant contributions to the systematic uncertainty: dependence on the $B\to X_c \ell \nu$ decay model, detector response, residual background, and $B_{reco}$ background subtraction.\\}}
\begin{tabular}{|c||c c c||c |c| c| c|}
\hline
\smallrule$P^*_{min}$&\multicolumn{7}{c|}{\mmx ~(GeV)} \\ 
($GeV$)& & stat. &sys.                  &$X_c$ Model & Detector & Background & $B_{reco}$ Background\\ \hline\hline
 0.9 &  2.0728& $\pm$0.0130& $\pm$0.0127&  0.0021  & 0.0092   & 0.0077     & 0.0037 \\ \hline
 1.0 &  2.0679& $\pm$0.0120& $\pm$0.0108&  0.0021  & 0.0088   & 0.0050     & 0.0030 \\ \hline
 1.1 &  2.0609& $\pm$0.0114& $\pm$0.0097&  0.0023  & 0.0087   & 0.0033     & 0.0016 \\ \hline
 1.2 &  2.0560& $\pm$0.0109& $\pm$0.0091&  0.0025  & 0.0083   & 0.0024     & 0.0013 \\ \hline
 1.3 &  2.0465& $\pm$0.0108& $\pm$0.0086&  0.0024  & 0.0080   & 0.0015     & 0.0013 \\ \hline
 1.4 &  2.0349& $\pm$0.0112& $\pm$0.0086&  0.0020  & 0.0081   & 0.0015     & 0.0017 \\ \hline
 1.5 &  2.0344& $\pm$0.0119& $\pm$0.0089&  0.0022  & 0.0081   & 0.0014     & 0.0026 \\ \hline
 1.6 &  2.0259& $\pm$0.0133& $\pm$0.0111&  0.0022  & 0.0102   & 0.0019     & 0.0035 \\ \hline
\end{tabular}
\end{center}
\end{table}

\begin{table}[ht]
\begin{center}
\caption[]{\label{tab:mom2} {\it Preliminary results for \mmxs\ for different values of $p^*_{min}$, with statistical and the total systematic errors. The last four columns show separately the four dominant contributions to the systematic uncertainty: dependence on the $B\to X_c \ell \nu$ decay model, detector response, residual background, and $B_{reco}$ background subtraction.\\}}
\vspace{-0.5cm}
\begin{tabular}{|c||c c c||c |c |c |c|}
\hline
\smallrule $P^*_{min}$ ~(GeV)&\multicolumn{7}{c|}{\mmxs ~($GeV^2$)} \\ 
& & stat. &  sys.& $X_c$ Model & Detector & Background & $B_{reco}$ Background\\ \hline\hline
 0.9 &  4.366& $\pm$0.049& $\pm$0.057& 0.009 & 0.034 & 0.039 & 0.023 \\ \hline
 1.0 &  4.338& $\pm$0.043& $\pm$0.046& 0.009 & 0.033 & 0.025 & 0.016 \\ \hline
 1.1 &  4.300& $\pm$0.040& $\pm$0.038& 0.010 & 0.032 & 0.016 & 0.006 \\ \hline
 1.2 &  4.276& $\pm$0.037& $\pm$0.035& 0.011 & 0.030 & 0.011 & 0.006 \\ \hline
 1.3 &  4.239& $\pm$0.036& $\pm$0.031& 0.010 & 0.028 & 0.007 & 0.006 \\ \hline
 1.4 &  4.190& $\pm$0.036& $\pm$0.029& 0.008 & 0.027 & 0.005 & 0.007 \\ \hline
 1.5 &  4.180& $\pm$0.038& $\pm$0.028& 0.008 & 0.026 & 0.005 & 0.006 \\ \hline
 1.6 &  4.146& $\pm$0.042& $\pm$0.034& 0.007 & 0.031 & 0.007 & 0.009 \\ \hline
\end{tabular}
\end{center}
\end{table}

\section{Systematic Uncertainties and Cross Checks}

\label{sec:Systematics}
Extensive studies have been performed to assess the systematic uncertainties in the measurement of \mmx\ and \mmxs\ as a function of $p^*_{min}$. 
The main sources of systematic errors are the subtraction of the combinatorical background of the $B_{reco}$ sample, the residual background estimate, inaccuracies in the modeling of the detector efficiency and particle reconstruction, and the impact of the modeling of the hadronic states on the efficiencies and mass measurements. The uncertainty related to the detector modeling and reconstruction is the dominant systematic error. Studies involve 
changes in the event selection and variations of the various corrections for particle reconstruction and resolutions, background suppression, and comparisons of results for various subsamples. The systematic errors are listed in Tables \ref{tab:mom1} and~\ref{tab:mom2}.

The combinatoric $B_{reco}$ background is the principal background source. It is determined from the $m_{ES}$ distribution as a function of $M_X$.  The  size of this background increases significantly for events with lower lepton momentum. The uncertainty in this background is estimated by varying the lower limit of the signal region in the $m_{ES}$ distribution.

The residual backgrounds are due to misidentified hadrons and non-prompt leptons from $\tau^{\pm}$ decays and from semileptonic decays of $D^{(*)}$ and $D_s$ mesons produced via $W^{\pm}$ fragmentation.  
The processes leading to ``right-sign'' background from $D^{(*)}$ and $D_s$ decays contribute primarily to higher masses $M_X$.  
These decay processes dominate the background uncertainties since the combined
uncertainty of their branching ratios is typically 30\%~\cite{PDG2002}.
Hadron misidentification contributes both ``right-sign" and ``wrong-sign" background. From a comparison of MC and data samples of events with ``wrong-sign" leptons we conclude that the MC estimate of the hadron misidentification agrees with the data, and we take the 5\% statistical error of this comparison as an estimate of the systematic error.

The uncertainties in the detector performance and particle reconstruction have been estimated by Monte Carlo simulations of the track and photon efficiencies.  Resolutions, fake rates, and background rates have been studied in detail by varying the adjustments to the MC simulation that are introduced to improve the agreement with data. These include in particular the requirements on $LAT$ and the minimum angular separation between a shower and the nearest track in the EMC.  
The root mean square (RMS) of the variation of the measured moments as a function of the lateral shower shape and the separation of the shower from the nearest charged track is assigned as the systematic error and is attributed to uncertainties in the simulation of hadronic interactions in the calorimeter. An error of 3\% in the energy response of the EMC was assumed to estimate the uncertainty due to the photon energy measurement.  The impact of track losses and fake tracks is largely independent of the minimum lepton momentum.   

The uncertainty in the acceptance variation is studied by varying all the branching ratios, in particular the composition of the hadronic high mass resonant and non-resonant states in the MC model.  The observed bias is very small and we assign the RMS of the bias as the systematic error of this correction.  The systematic error due to these variations is roughly constant for all \psm.

\begin{figure}[t]
\begin{center}
\begin{picture}(15.,7.)
\put(8.3,0.0){\mbox{\epsfig{file=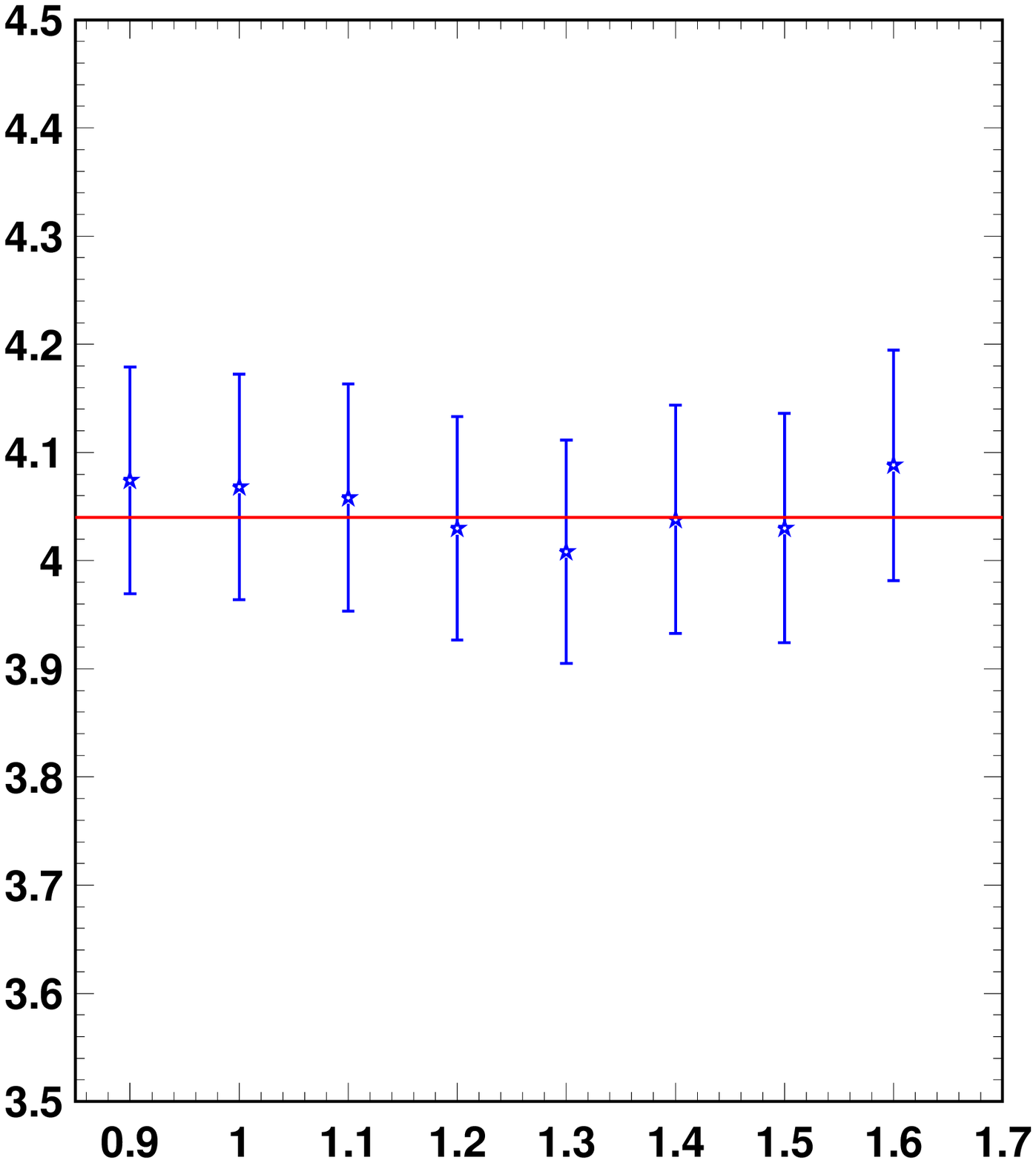, scale=0.33}}}
\put(0.0,0.0){\mbox{\epsfig{file=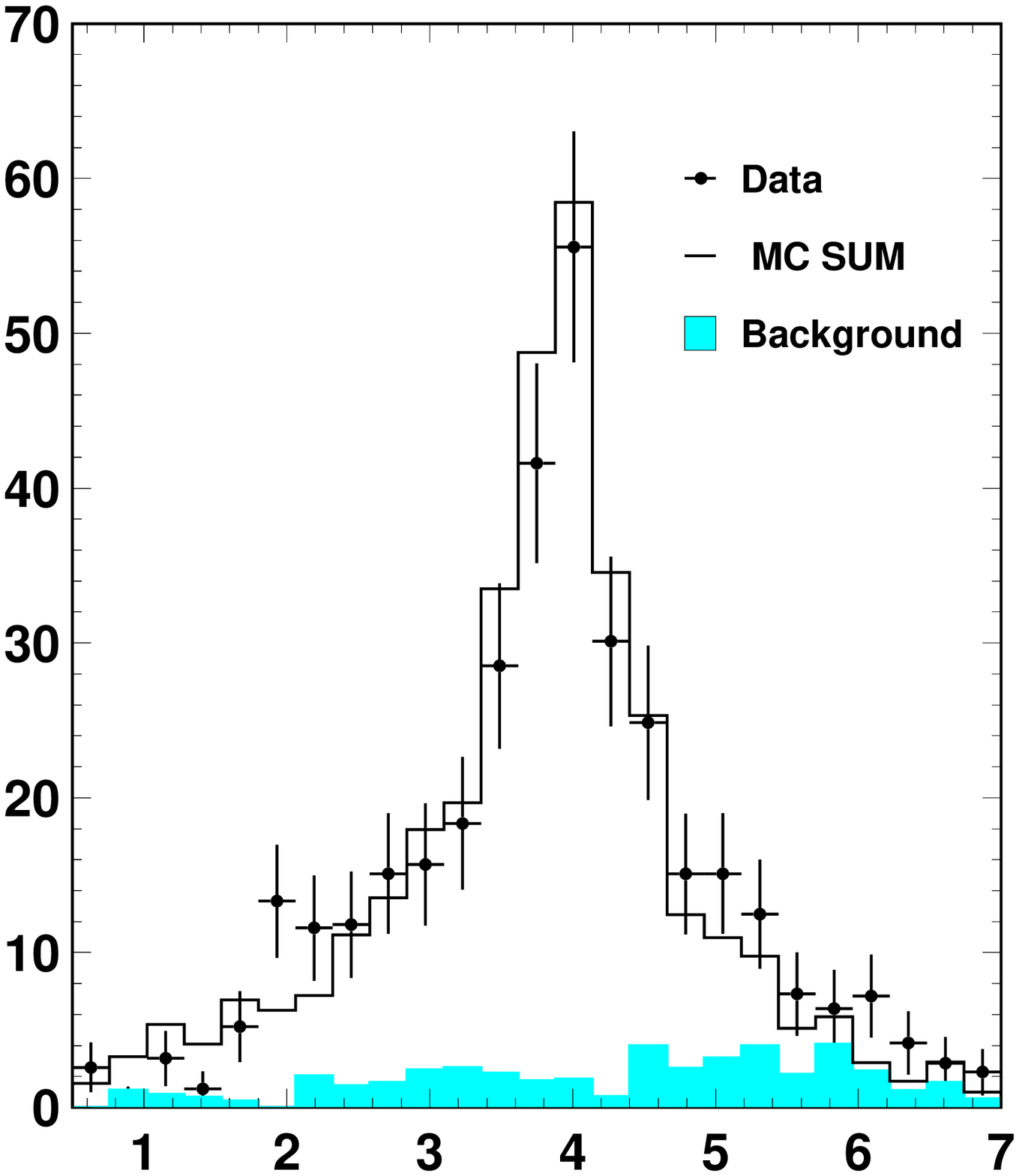, scale=0.33}}}
\put(8.3,4.2){\mbox{\begin{turn}{90}{$\langle M_X^2 \rangle$ ($\gev^2$)}\end{turn}}}
\put(4.1,-0.1){$M_X^2$ ($\gev^2$)}
\put(0.1,4.3){\mbox{\begin{turn}{90}{$N/ 260 \mev^2 $}\end{turn}}}
\put(12.3,0.0){$\psm$ ($\gev$)}
\end{picture}
\end{center}
\vspace{-0.5cm}
\caption{\em \label{partdstar} Verification of the measurement of mass distribution  and \mmxs\ moment for $B \ra D^* \ell \nu$.  Left: The data (points) are compared to the MC prediction (histogram) for all selected events. The MC estimated background is shown as the shaded area at the bottom.  Right: The measured \mmxs\ moments (data points) as a function of $p^*_{min}$ are compared to the expectation, $M_{D^*}^2$, marked as a solid line.}
\end{figure}

The uncertainty in the $M_X$ calibration is estimated by replacing the nine calibration curves by a single one.  The observed change in the moments is assigned as the error in the $M_X$ calibration.  
To verify the accuracy of the mass calibation method we apply it  to a sample of $B^0 \ra D^{*+} \ell^- \nu$ decays that are identified by partial reconstruction of the decay $D^{*+} \to D^0 \pi^+$ via the identification of the low momentum $\pi^+$.
Figure~\ref{partdstar} shows the hadronic mass-squared distribution for these events together with the measured \mmxs\ moments which agree well with the expectation to within the statistical errors and show a very minor dependence on \psm. 

The calibration method has also been extensively validated on MC samples.
Since the moment measurements for different lepton momentum thresholds \psm\ are highly correlated, the moments were determined from the measured $M_X$ and $M_X^2$ distributions for four separate intervals in the lepton momentum and compared with the true moments of MC generated distributions. The measured values agree well with the true values.

\begin{figure}[H]
\begin{center}
\begin{picture}(15.,9.)
\put(-1.8,0.0){\mbox{\epsfig{file=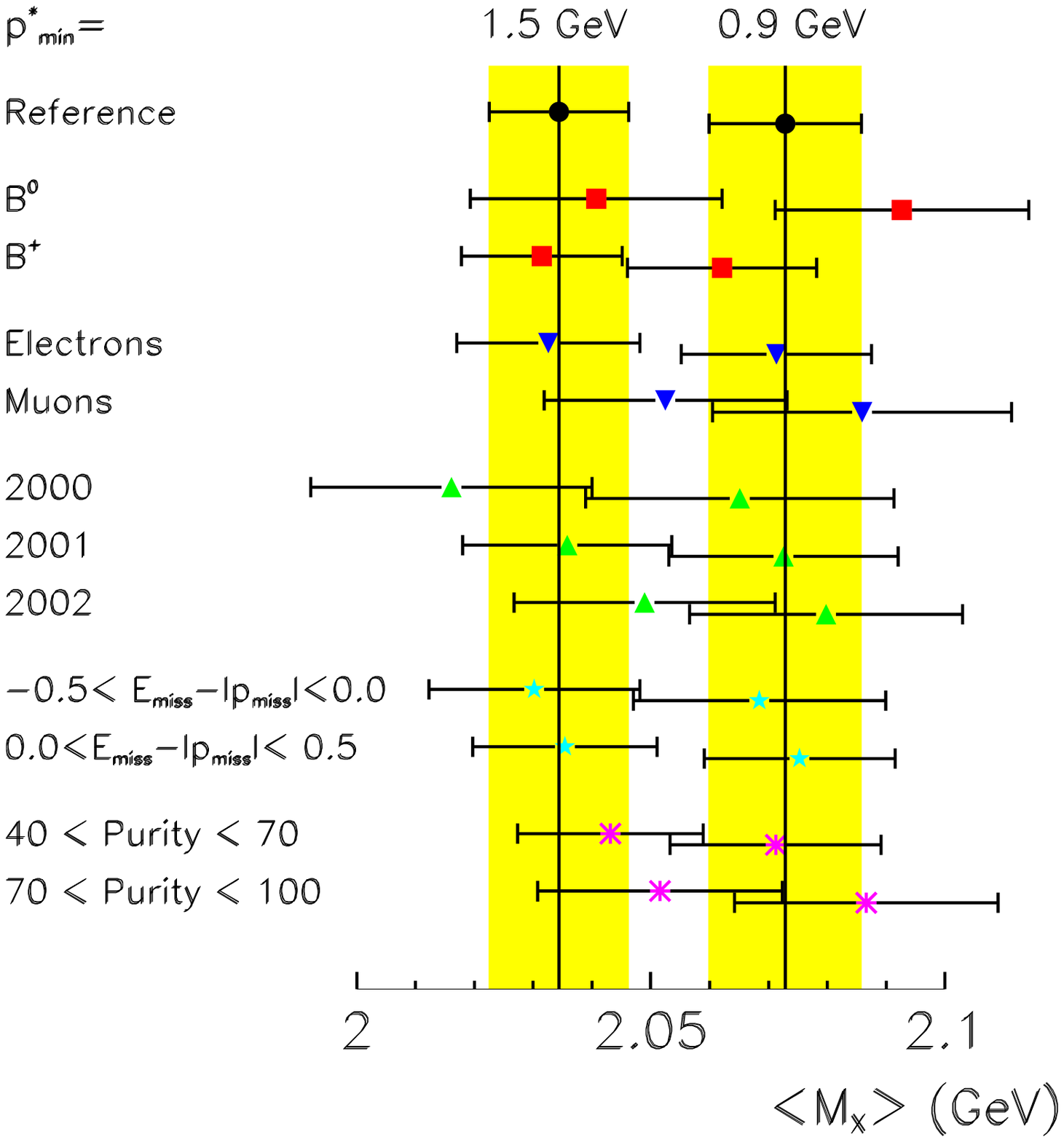, scale=0.5}}}
\put(6.7,0.0){\mbox{\epsfig{file=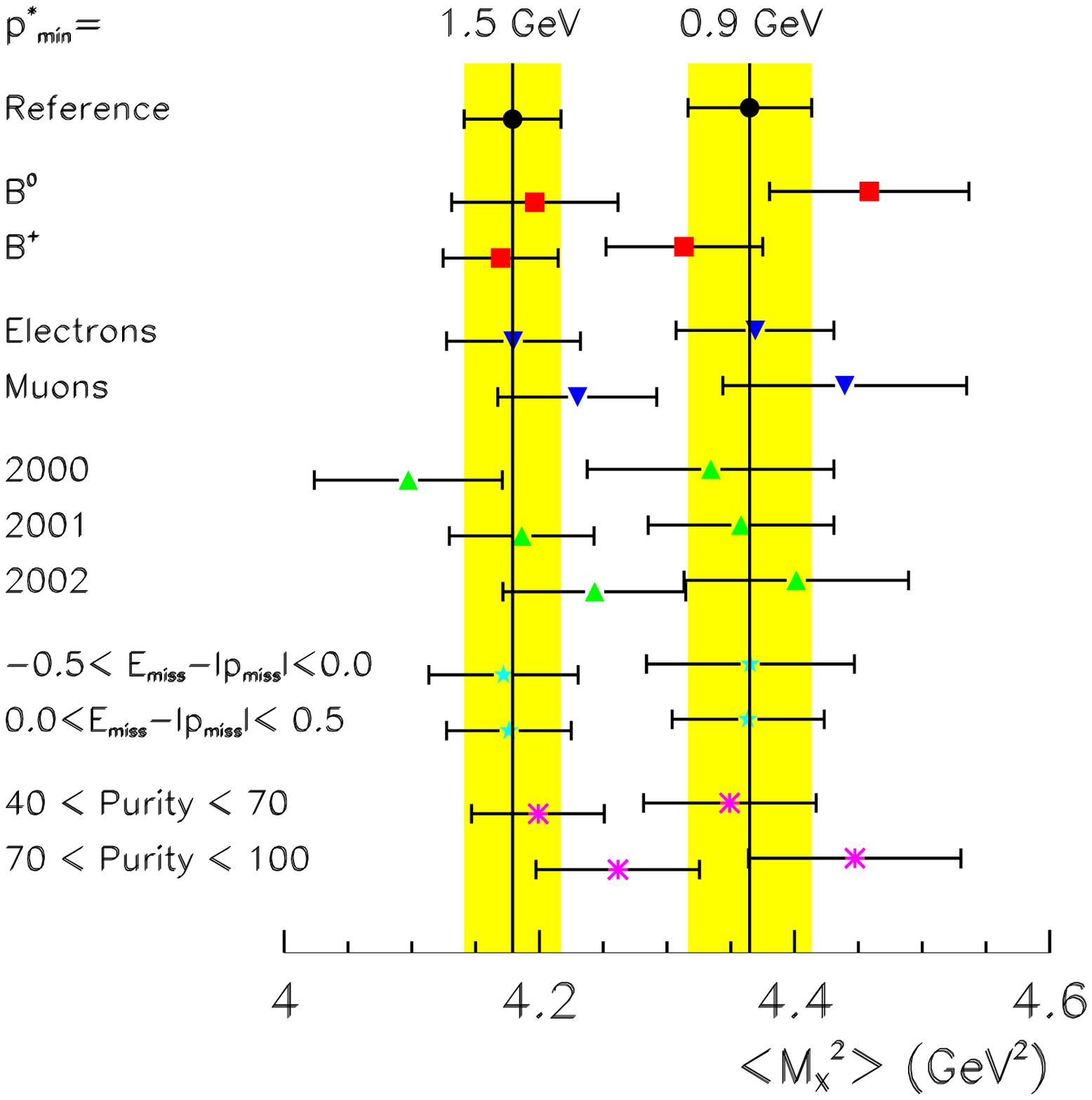, scale=0.5}}}
\end{picture}   
\end{center}
\vspace{-0.5cm}
\caption{\em \label{ursl} A comparison of the results for \mmx\ and \mmxs\ obtained from pairwise independent subsamples (statistical errors only).}
\end{figure}

A variety of tests are performed to test the stability of the moment measurements.  
The data are divided into several independent subsamples and the whole analysis is applied separately to each of them, including \breco\ background subtraction and mass calibration.
The consistency of the measurement is
checked for
$B^{\pm}$ and $B^0$, decays to electrons and muons, events from different run periods, with different $E_{miss} - |\vec{p}_{miss}|$, and of different \breco\ sample purity ${\cal P}$. The results obtained from the two $E_{miss} - |\vec{p}_{miss}|$ regions specifically address our understanding of the detector performance as this variable is very sensitive to the reconstruction of particles. A comparison of these results is shown in Figure~\ref{ursl}.  No significant systematic variations are observed.

\section{Conclusions and Interpretation of Results}
\label{sec:Physics}

We have performed a preliminary measurement of the first and second moment 
$\langle M_X\rangle$ and $\langle M_X^2\rangle$
of the hadronic mass distribution in semileptonic $B$ decays.  
These measurements have been carried out as a function of the threshold lepton momentum, \psm, extending from $0.9 \gev$ to $1.6 \gev$. 
They are sensitive to the production of higher-mass charm states, 
both resonant and non-resonant. 
The reconstruction of a hadronic decay of one of the two $B$ mesons and the 
kinematic fit to the full event have resulted in measurements with
comparable statistical and systematic uncertainties. 
In addition, we have evaluated and taken into account the correlation between the individual data points.

\begin{figure}[H]
\begin{center}
\begin{picture}(15.,7.5)
\put(-1.0,0.0){\mbox{\epsfig{file=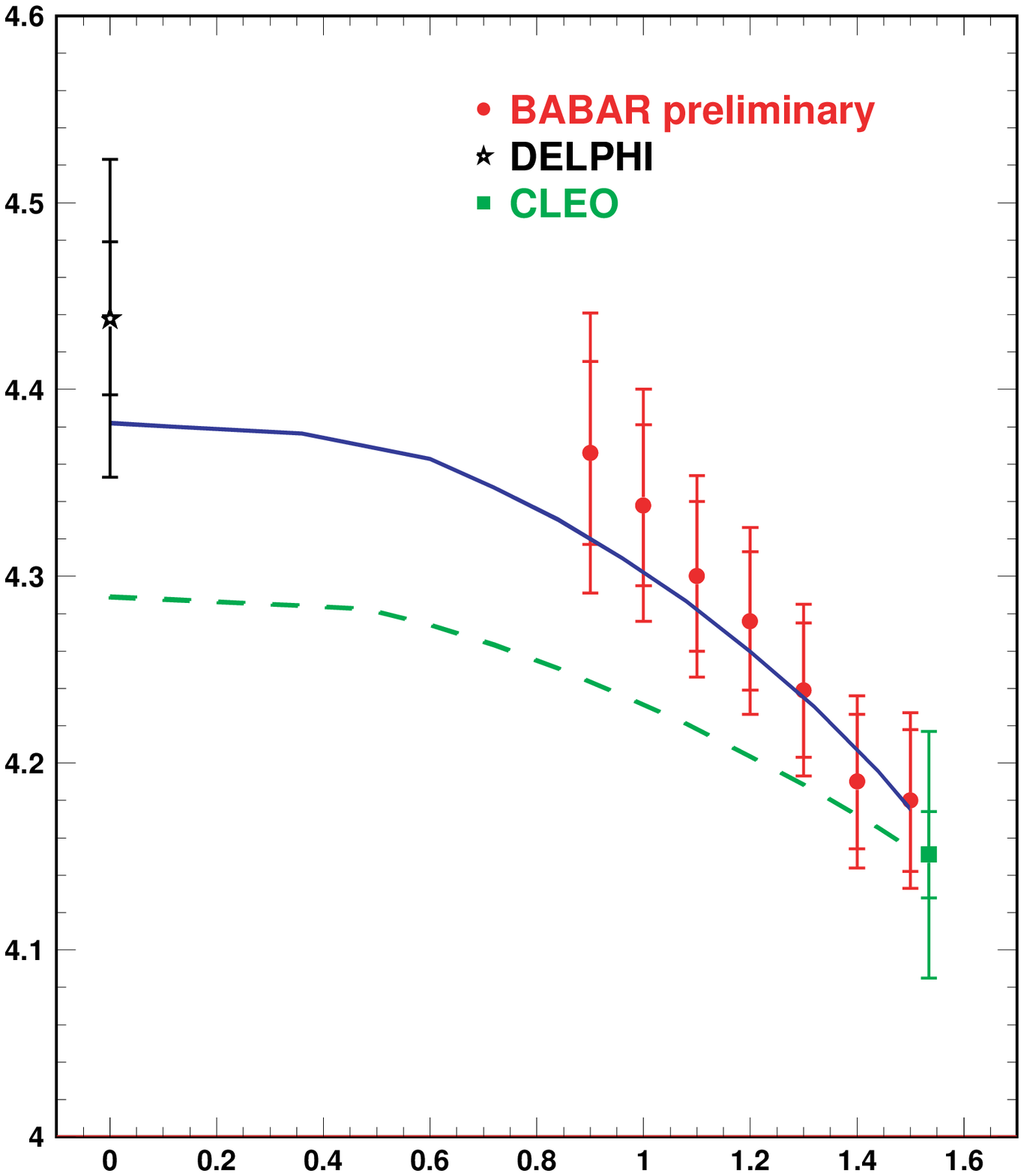,height=9.0cm,width=9.0cm}}}
\put(7.5,0.0){\mbox{\epsfig{file=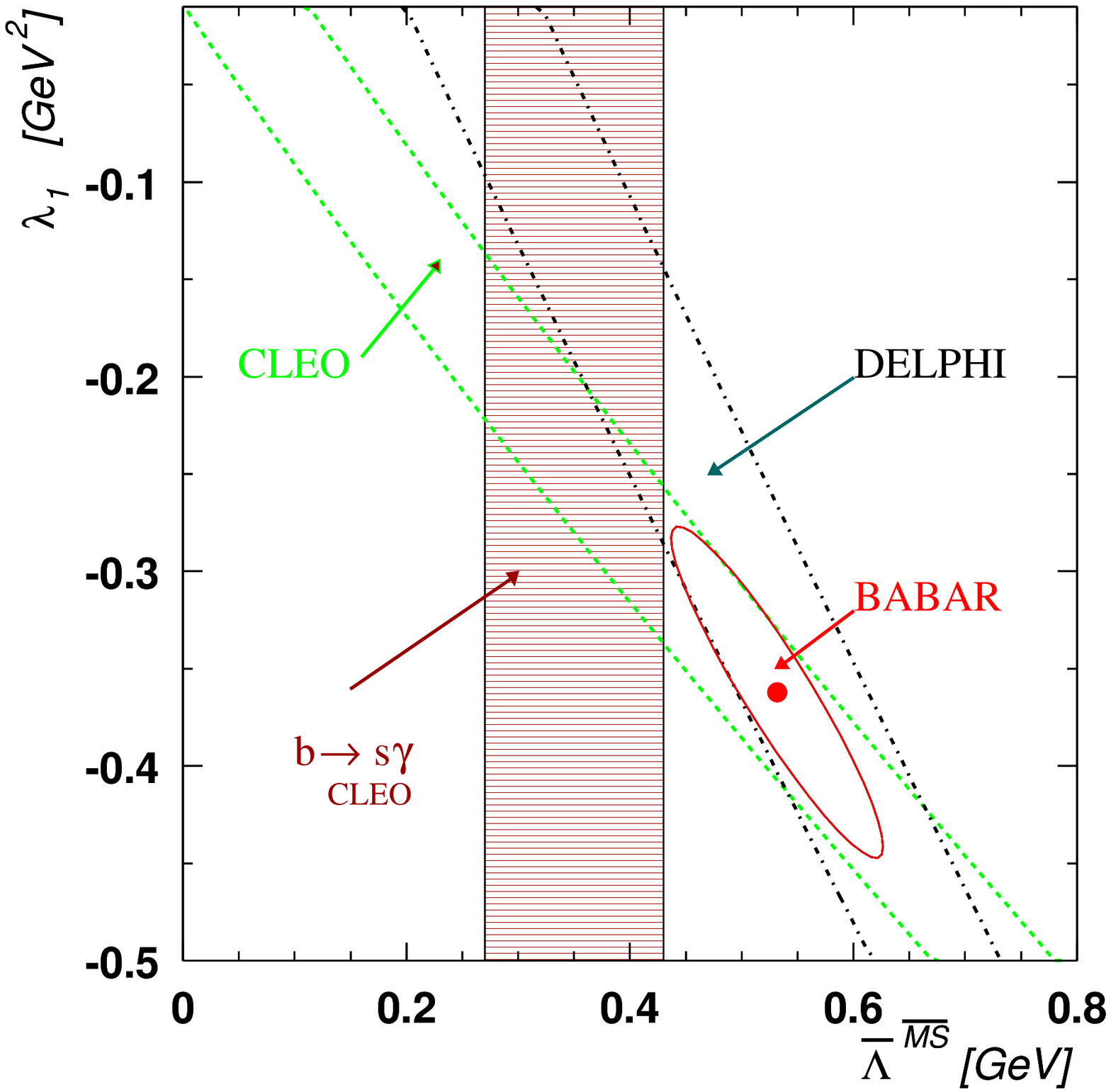,scale=0.45}}}
\put(0.6,7.4){a)}
\put(14.8,7.4){b)}
\put(5.3,0.1){$p^*_{min} [\gev]$}
\put(-0.6,5.7){\begin{turn}{90}\mmxs $[\gev^2]$\end{turn}}
\end{picture}   
\end{center}
\vspace{-0.8cm}
\caption{\em 
Comparison of results obtained from hadronic mass moments from the \babar~, CLEO and DELPHI experiments: a)  the measured hadronic mass moments \mmxs\ as a function of the lepton momentum threshold \psm.  The solid curve is a fit to the \babar\ data; the dashed curve is the HQE prediction obtained from measurements by CLEO of $\langle E_{\gamma} \rangle$  and \mmxs\ for \psm = 1.5 \gev.
b) Constraints on the HQE parameters $\bar{\Lambda}^{\overline{MS}}$ and $\lambda_1$ from: the fit to the \babar\ \mmxs\ measurements ($\Delta \chi^2=1$ ellipse), the CLEO and DELPHI \mmxs\ measurements (one-$\sigma$ bands), and the  CLEO $b\ra s\gamma$ energy moment (one-$\sigma$ band). The constraints are derived in the $\overline{MS}$ scheme to ${\cal O}(1/m_B^3)$.
\label{fig:lbarl1} }
\begin{center}
\begin{picture}(15.,8.6)
\put(-1.0,0.0){\mbox{\epsfig{file=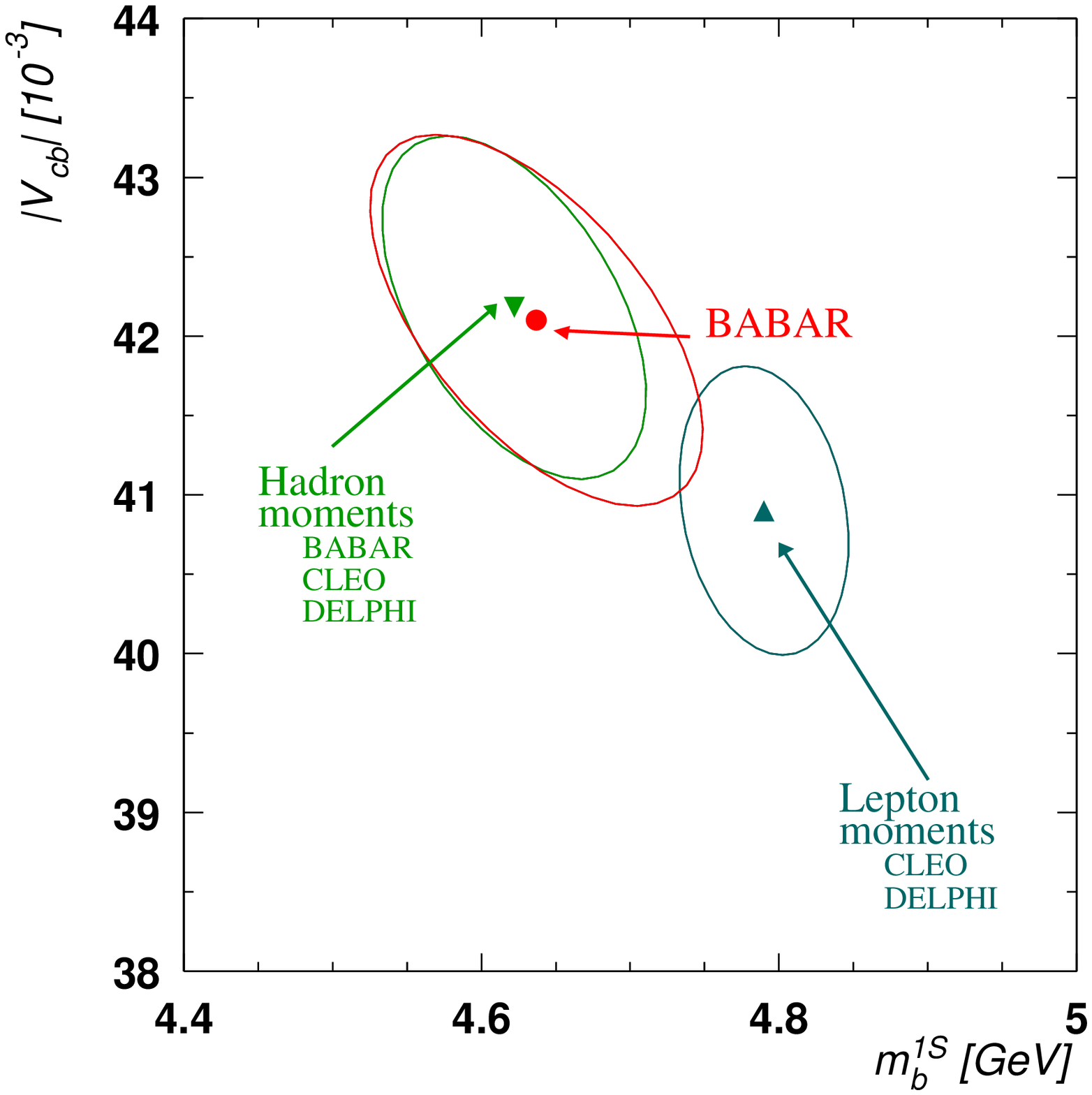,scale=0.45}}}
\put(7.5,0.0){\mbox{\epsfig{file=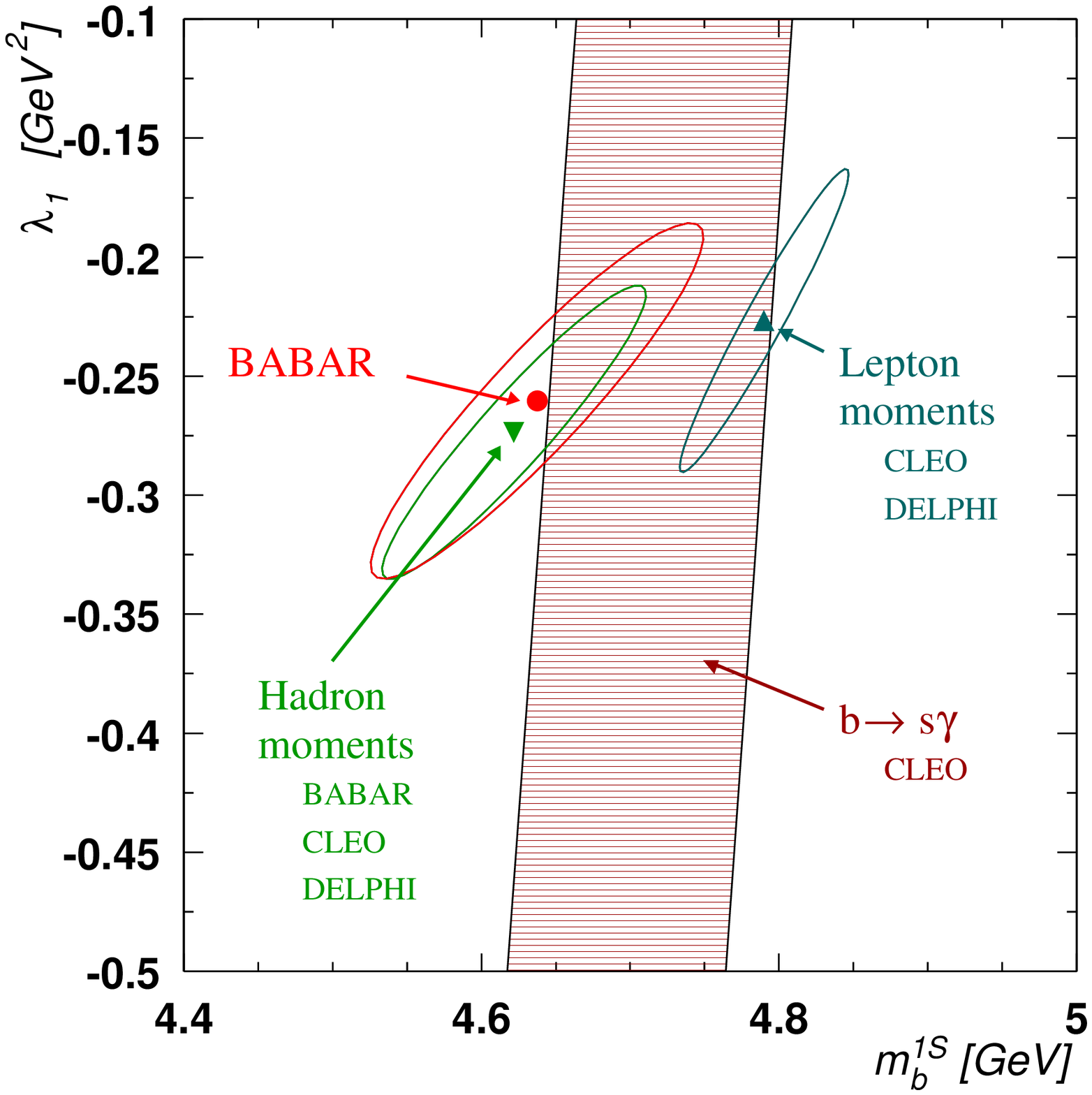,scale=0.45}}}
\put(0.6,7.4){a)}
\put(14.8,7.4){b)}
\end{picture}   
\end{center}
\vspace{-0.8cm}
\caption{\em Constraints on HQE parameters obtained from hadronic moments as measured by \babar, all hadronic moments combined (\babar, CLEO and DELPHI), and the combined lepton moments (CLEO and DELPHI),
projected a) in the ($|V_{cb}|$, $m_b^{1S}$) plane and b) the ($\lambda_1$, $m_b^{1S}$) plane.
Also indicated is the constraint from the $b\ra s\gamma$ photon energy moment measured by CLEO.
\label{fig:vcbmb} }
\end{figure}

Figure \ref{fig:lbarl1}a shows the moments $\langle M_X^2 \rangle$ 
as a function of \psm. As a result of several changes to the analysis and data selection, these new measurements fall 
below the preliminary measurements presented previously~\cite{Aubert:2002pm}. 
 
If this change were of purely statistical origin the level of compatibility of the two measurements
would be approximately 5\%.
However, there is evidence that the systematic uncertainty related to the track and neutral selection used in~\cite{Aubert:2002pm} was underestimated.
Since then, the total \BB\ sample has doubled and the event selection, in particular the rejection of background tracks and photons, has been significantly improved. Furthermore, the moments are now extracted directly 
from the $M_X^2$ distribution, resulting in measurements that are much less sensitive to the uncertainties in the branching ratios and mass distributions of the individual hadronic states. 

The hadronic mass moments can be calculated in the framework of Heavy Quark Expansion (HQE), which expresses perturbative and non-perturbative corrections to the parton level calculations in powers of the strong coupling constant, $\alpha_s(m_b)$, and in inverse powers of the $B$ meson mass, $1/m_B$. 
We derive the two leading HQE parameters, $\lambda_1$ and $\bar{\Lambda}$, that determine the non-perturbative corrections at ${\cal O}(1/m_B^2)$ from a minimum $\chi^2$ fit of the \mmxs\ moments measured for seven different lepton momentum thresholds. In these fits we take into account the strong correlations between the measured moments for different $p^*_{min}$.  In these expansions we use the coefficients calculated  in the $\overline{MS}$ 
mass scheme~\cite{Falk:1998jq}. We use $\lambda_2=0.128\pm0.010~\gev^2$
as determined from the $B - B^*$ and $D - D^*$ mass splitting~\cite{Gremm:1997df}. The six non-perturbative parameters 
$\rho_{1,2}$ and ${\cal T}_{1,2,3,4}$ that appear at order ${\cal O}(1/m_B^3)$, 
are expected to be of the order $\Lambda^3_{QCD}$ and from dimensional arguments are assumed to be smaller than $(0.5)^3 \gev^3$~\cite{Gremm:1997df}.  For this fit, we set their values to ${\cal T}_i=0.0 \gev^3$, $\rho_{1}=\frac{1}{2}(0.5)^3 \gev^3$, and 
we eliminate $\rho_2$ by making use of the relation between $\rho_2, {\cal T}_2$ and ${\cal T}_4$ and the $D^*-D$ and $B^*-B$ mass splittings represented in~\cite{Gremm:1997df}.
The fitted parameters result in a good description of the 
\babar\ data, as indicated by the solid line in Figure~\ref{fig:lbarl1}. The fit result also agrees well with the measurement by the CLEO
Collaboration~\cite{Cronin-Hennessy:2001fk} for lepton momenta above 1.5 \gev and is consistent with the
DELPHI Collaboration~\cite{Calvi:2002wc} measurement corresponding to $\psm = 0 \gev$ (see Table~\ref{tab:extmom}).

If we combine the first moment of the $b \ra s\gamma$ photon energy 
spectrum measured by CLEO~\cite{Chen:2001fj}, $\langle E_\gamma \rangle =   2.346 \pm 0.034 ~\gev$,
with the CLEO \mmxs\ measurement for  $\psm =1.5 \gev$
and calculate $\lambda_1$ and $\bar{\Lambda}$, the result does not give a consistent description of all available \mmxs\ measurements (see dashed line in Figure \ref{fig:lbarl1}a). Here we used the coefficients for the HQE of $\langle E_{\gamma} \rangle$ from~\cite{Bauer:2002sh}.
The same effect is also illustrated in Figure~\ref{fig:lbarl1}b,
where the result of the fit  to the \babar\  $\langle M_X^2 \rangle$ measurements is displayed in form of a $\Delta\chi^2=1$ contour (39\% confidence level) in the ($\lambda_1~,~\bar{\Lambda}^{\overline{MS}}$) plane. 
The resulting values for the HQE parameters are
$\bar{\Lambda}^{\overline{MS}}=0.53\pm0.09~\gev$ and
$\lambda_1=-0.36\pm0.09~\gev^2$.
These compare to earlier CLEO measurements of $\bar{\Lambda}^{\overline{MS}}=0.35\pm0.07~\gev$ and
$\lambda_1=-0.24\pm0.07~\gev^2$~\cite{Cronin-Hennessy:2001fk}.
In Figure~\ref{fig:lbarl1}b,
we also show the constraints from the hadronic mass moment measurements by CLEO and DELPHI, and from the $\langle E_\gamma  \rangle$ moment from CLEO.
The results of the $\langle M_X^2 \rangle$ measurements are in good agreement.
It should be emphasized that the one-$\sigma$ areas indicated here  
do not include uncertainties due to the order ${\cal O}(1/m_B^3)$  terms.\\

Since the HQEs for the second hadronic mass moments are analogues to the total semileptonic rate, and since they contain the same non-perturbative parameters we can combine the seven measured moments with the semileptonic decay rate to perform a more comprehensive test of the HQE and to extract $|V_{cb}|$ and the $b$-quark mass, $m_b$.  For this extraction we  choose the $1S$ mass scheme as recommended in~\cite{Bauer:2002sh} because it leads to a better 
convergence of the perturbative expansion than the $\overline{MS}$ scheme.
In this scheme, $\bar{\Lambda}^{1S}$ is related to the $b$-quark mass via
$\bar{\Lambda}^{1S}=m_{\Upsilon}/2 - m_b^{1S}$.  The coefficients for the expansions are taken from Bauer $et~al.$~\cite{Bauer:2002sh}.
We take into account the uncertainties of the other parameters of the expansion, i.e., 
for the independent ${\cal O}(1/m_B^3)$ matrix elements 
we equate $\rho_1=\frac{1}{2}(0.5)^3\pm\frac{1}{2}(0.5)^3 \gev^3$, ${\cal T}_i=0.0\pm (0.5)^3 \gev^3$, 
and make the same assumption for $\rho_2$ as stated above.
For the semileptonic decay width,  we take $\Gamma_{SL} = (4.37 \pm 0.18) \times 10^{-11}\mev$, as derived from individual 
\babar\ measurements (see Table~\ref{tab:gsl}).

\begin{table}[t]
\begin{center}
\caption[]{\label{tab:extmom} {\it Moment measurements from other experiments that are used in the fits.
\\}}
\vspace{0.3cm}
\begin{tabular}{|c|c c|}
\hline\hline
Experiment & Quantity & Value\\ \hline
                            &  &Hadron Mass Moments   \\ \hline
CLEO~\cite{Cronin-Hennessy:2001fk}    &  \mmxs  &  $4.152 \pm 0.066 \gev^2$\\
DELPHI~\cite{Calvi:2002wc}  &  \mmxs  &  $4.430 \pm 0.080 \gev^2$\\ \hline\hline
                            &  &Lepton Energy Moments   \\ \hline
CLEO~\cite{Briere:2002hw}   & $R_0^e$  & $0.6184 \pm 0.0023\gev$ \\
CLEO~\cite{Briere:2002hw}   & $R_1^e$  & $1.7817 \pm 0.0013 \gev$  \\
CLEO~\cite{Briere:2002hw}   & $R_0^{\mu}$  & $0.6189 \pm 0.0030 \gev$ \\ 
CLEO~\cite{Briere:2002hw}   & $R_1^{\mu}$  & $1.7802 \pm 0.0016 \gev$ \\
DELPHI~\cite{Calvi:2002wc}  & $R_1$  & $1.383 \pm 0.014 \gev$ \\ 
\hline\hline
\end{tabular}
\end{center}
\begin{center}
\caption[]{\label{tab:gsl} {\it Measured quantities used in the determination of $\Gamma_{SL}$.\\}}
\vspace{0.3cm}
\begin{tabular}{|c|c c|}
\hline
Experiment & Quantity & Value\\ \hline
\babar\ ~\cite{Aubert:2002sh}  & $\tau_{\Bz}$  & $1.523 \pm0.033~ps$\\
\babar\ ~\cite{Aubert:2002gi} &  $\tau_{\Bz}$ & $1.529 \pm0.031~ps$\\ 
\babar\ ~\cite{Aubert:2001uw} & $\tau_{\Bz}$  & $1.546 \pm0.039~ps$\\
\babar\ ~\cite{Aubert:2001uw} & $\tau_{\Bp}$ &  $1.673 \pm 0.039 ~ps$ \\
HFAG~\footnotemark[1] & $f_+/f_0$  &  $1.053 \pm 0.055$\\ 
\babar\ ~\cite{Aubert:2002uf} &  ${\cal B}(B \to X \ell \nu)$ & $ 0.1087\pm0.0035$\\
\babar\ ~\cite{delRe:2003if} &  ${\cal B}(B\to X_u \ell\nu)$ & $(2.14\pm0.54) \times 10^{-3}$ \\ \hline
\end{tabular}
\end{center}
\end{table}
\footnotetext[1]{Heavy Flavor Averaging Group: $f_+/f_0$  from~\cite{PDG2002} has been rescaled by the lifetime ratio of \Bp to \Bz from~\cite{PDG2003}.}

Using the HQE expansions for the moments \mmxs\ and $\Gamma_{SL}$ we determine three parameters by a $\chi^2$ minimization:
$|V_{cb}|$, $m_{b}^{1S}$, and $\lambda_{1}$.  
The fit takes into account the experimental errors and their correlations.  For treatment of the theoretical errors on the perturbative terms we follow the suggestions by Bauer $et~al.$~\cite{Bauer:2002sh} to include errors estimated from dimensional analysis ($dim$) and from the size of the BLM term in the perturbative series~\cite{Brodsky:1983gc}. These errors are added in quadrature and assumed to be 100\% correlated for the various expansions.   In addition, the sensitivity to the  $1/m_B^3$ terms is estimated from the change in the results of the $\chi^2$ minimization for a variation of the $1/m_B^3$ matrix elements, assuming a flat, rather than Gaussian, error distribution.
From these fits to the HQE expansions ~\cite{Bauer:2002sh} for the \babar\ hadronic moments and the semileptonic decay rate, based on \babar\ measurements alone, we obtain
\begin{eqnarray}
 |V_{cb}|   &=& (42.10 \pm 1.04(exp)  \pm 0.52(dim \oplus BLM) \pm 0.50(1/m_B^3)) \times 10^{-3}, \nonumber \\ 
 m_b^{1S}  &=&4.638 \pm 0.094(exp) \pm 0.062(dim \oplus BLM) \pm 0.065(1/m_B^3)\gev, \quad \mbox{and} \nonumber \\
 \lambda_1  &=&-0.26 \pm 0.06(exp) \pm 0.04(dim \oplus BLM) \pm 0.04(1/m_B^3)\gev^2. 
\quad \nonumber
\end{eqnarray}
These results are in good agreement with other determinations~\cite{Briere:2002hw,Aubert:2002uf,Battaglia:2002tm}. 
The correlation between $\lambda_1$ and $m_b^{1S}$ is 0.92, between \Vcb\ and $m_b^{1S}$ and $\lambda_1$ is $-0.60$ and $0.51$, respectively.

The shape of the lepton energy spectrum can be sampled in terms of moments that are also predicted by HQE calculations.  Thus, to further test the HQE calculations 
we perform a separate fit using the measured lepton energy moments, as listed in Table~\ref{tab:extmom}, and the semileptonic rate $\Gamma_{SL}$.  
Again, we are relying on the calculations of the expansion coefficients by Bauer $et~al.$~\cite{Bauer:2002sh}.
Since the $\chi^2$ minimization requires a proper treatment of the correlations among the measurements, we only consider published 
results that are either uncorrelated with all other measurements 
or for which the full covariance matrix is available.  Thus
we select four lepton moment measurements by the CLEO Collaboration~\cite{Briere:2002hw}
and one measurement by the DELPHI Collaboration~\cite{Calvi:2002wc}.
In Figure~\ref{fig:vcbmb}(a,b) the fit results are shown separately for hadron and lepton energy moments, including the $\Delta\chi^2=1$ contours in the ($|V_{cb}|~,~m_b^{1S}$) and ($\lambda_1~,~m_b^{1S}$) planes, taking into account experimental and estimated theoretical errors ($dim \oplus BLM$ only) as explained above.
The fit contours of hadronic and lepton energy moments do not overlap in either of the planes, but they each overlap with the one-$\sigma$ constraint from the photon energy moments measured by CLEO. 

In conclusion, the  measurements of the hadronic mass moments in semileptonic $B$ decays presented here agree well with measurements by other experiments of the same quantity, but have been extracted by a new technique which reduces the model dependence. By combining the second moments of the hadron mass distributions with earlier \babar\ measurements of the semileptonic decay rate, we have been able to extract 
the HQE parameters $m_b^{1S}$ and $\lambda_1$ and reduce the theoretical uncertainties on \vcb . To further improve these measurements and examine the consistency of these results with measurements of other inclusive distributions in the context of HQE calculations,  it will be very important to extend the currently available measurements to higher precision and  thereby provide more stringent constraints on the ${\cal O}(1/m_B^3)$  parameters.

\unitlength1.0cm 

\section{Acknowledgments}

\label{sec:Acknowledgments}


The authors wish to express their appreciation to C. W. Bauer, Z. Ligeti, M. Luke and A. V. Manohar for providing the coefficients for the HQE calculations and
for their input to the development of the fitting software for the extraction of the 
HQE parameters from the moment measurements.
They would like to thank them as well as N.G. Uraltsev for valuable discussions.


We are grateful for the 
extraordinary contributions of our \pep2\ colleagues in
achieving the excellent luminosity and machine conditions
that have made this work possible.
The success of this project also relies critically on the 
expertise and dedication of the computing organizations that 
support \babar.
The collaborating institutions wish to thank 
SLAC for its support and the kind hospitality extended to them. 
This work is supported by the
US Department of Energy
and National Science Foundation, the
Natural Sciences and Engineering Research Council (Canada),
Institute of High Energy Physics (China), the
Commissariat \`a l'Energie Atomique and
Institut National de Physique Nucl\'eaire et de Physique des Particules
(France), the
Bundesministerium f\"ur Bildung und Forschung and
Deutsche Forschungsgemeinschaft
(Germany), the
Istituto Nazionale di Fisica Nucleare (Italy),
the Foundation for Fundamental Research on Matter (The Netherlands),
the Research Council of Norway, the
Ministry of Science and Technology of the Russian Federation, and the
Particle Physics and Astronomy Research Council (United Kingdom). 
Individuals have received support from 
the A. P. Sloan Foundation, 
the Research Corporation,
and the Alexander von Humboldt Foundation.

\bibliography{refs}         

\bibliographystyle{h-physrev2-original}   %

\end{document}